\documentclass[sigconf]{templates/acmart/acmart}
\AtBeginDocument{%
  }

\copyrightyear{2024}
\acmYear{2024}
\setcopyright{rightsretained}
\acmConference[CCS '24]{Proceedings of the 2024 ACM SIGSAC Conference on Computer and Communications Security}{October 14--18, 2024}{Salt Lake City, UT, USA}
\acmBooktitle{Proceedings of the 2024 ACM SIGSAC Conference on Computer and Communications Security (CCS '24), October 14--18, 2024, Salt Lake City, UT, USA}
\acmDOI{10.1145/3658644.3690235}
\acmISBN{979-8-4007-0636-3/24/10}

\usepackage{graphicx}
\usepackage{xcolor}

\usepackage{here}
\usepackage{float}

\usepackage{hyperref}

\usepackage{tabularx}
\newcolumntype{Y}{>{\centering\arraybackslash}X}
\newcolumntype{L}[1]{>{\raggedright\let\newline\\\arraybackslash\hspace{0pt}}m{#1}}
\newcolumntype{C}[1]{>{\centering\let\newline\\\arraybackslash\hspace{0pt}}m{#1}}
\newcolumntype{R}[1]{>{\raggedleft\let\newline\\\arraybackslash\hspace{0pt}}m{#1}}

\usepackage{multirow}
\usepackage{booktabs}

\usepackage{subcaption}

\usepackage{comment}

\usepackage{adjustbox}

\usepackage{algorithm}
\usepackage[noend]{algpseudocode}
\algrenewcommand\algorithmicrequire{\textbf{\ \ Input:}}
\algrenewcommand\algorithmicensure{\textbf{Output:}}

\usepackage{amsfonts}
\usepackage{amsmath}

\usepackage{siunitx}

\usepackage{enumitem}

\usepackage{svg}
\usepackage{calc}

\usepackage{csvsimple,booktabs}

\usepackage{tikz}
\usetikzlibrary{matrix,calc,shapes}
\usetikzlibrary{matrix}
\usepackage{pgfplots}
\pgfplotsset{compat=1.18}

\newcommand{\mcrot}[4]{\multicolumn{#1}{#2}{\rlap{\rotatebox{#3}{#4}~}}}%

\usepackage[]{mdframed}

\definecolor{LightGray}{gray}{0.9}
\definecolor{DarkGray}{gray}{0.1}

\usepackage{xspace}

\newlist{questions}{enumerate}{1}
\setlist[questions,1]{label={\textbf{RQ\arabic*:}},ref={RQ\arabic*},left=-4pt,labelsep=0.5em,itemindent=2em,listparindent=\parindent}

\newlist{challenges}{enumerate}{1}
\setlist[challenges,1]{label={\textbf{C\arabic*:}},ref={C\arabic*},left=0pt,labelsep=0.5em,itemindent=1em,listparindent=\parindent}

\usepackage[nolist]{acronym}

\begin{acronym}
\acro{AES}{Advanced Encryption Standard}
\acro{AI}{artificial intelligence}
\acro{ALU}{arithmetic logic unit}
\acro{AOI}{and-or-invert}
\acro{API}{application programming interface}
\acro{ARX}{add-rotate-XOR}
\acro{ASIC}{application-specific integrated circuit}
\acro{ASIP}{application specific instruction-set processor}
\acro{AS}{active serial}

\acro{BFS}{breadth-first search}
\acro{BGA}{ball grid array}
\acro{BRAM}{block-\acs{RAM}}

\acro{CFG}{control flow graph}
\acro{CLB}{configurable logic block}
\acro{CPU}{central processing unit}

\acro{DC}{direct current}
\acro{DDR}{double data rate}
\acro{DFA}{differential frequency analysis}
\acro{DFT}{discrete Fourier transform}
\acro{DSO}{digital storage oscilloscope}
\acro{DSP}{digital signal processing}
\acro{DUT}{device under test}

\acro{EDA}{electronic design automation}
\acro{EEPROM}{electrically erasable programmable \acs{ROM}}
\acro{EMA}{electro-magnetic emanation}
\acro{EM}{electro-magnetic}

\acro{FFT}{fast Fourier transformation}
\acro{FF}{flip-flop}
\acro{FIB}{focused ion beam}
\acro{FIFO}{first in first out}
\acro{FIR}{finite impulse response}
\acro{FPGA}{field-programmable gate array}
\acro{FSM}{finite state machine}

\acro{GNN}{graph neural network}
\acro{GPIO}{general purpose input/output}
\acro{GUI}{graphical user interface}

\acro{HDL}{hardware description language}
\acro{HD}{Hamming distance}
\acro{HF}{high frequency}
\acro{HRE}{hardware \acl{RE}}
\acro{HSM}{hardware security module}
\acro{HW}{Hamming weight}

\acro{I2C}[I\textsuperscript{2}C]{Inter-Integrated Circuit}
\acro{I2S}[I\textsuperscript{2}S]{Inter-IC Sound}
\acro{IC}{integrated circuit}
\acro{ICAP}{Internal Configuration Access Port}
\acro{IIR}{infinite impulse response}
\acro{IO}[I/O]{input/output}
\acroplural{IO}[I/Os]{inputs/outputs}
\acro{IOB}{Input Output Block}
\acro{IOT}[IoT]{Internet of things}
\acro{IP}{intellectual property}
\acro{IPSW}{iPod Software}
\acro{ISA}{instruction set architecture}
\acro{IV}{initialization vector}

\acro{JTAG}{Joint Test Action Group}

\acro{LFSR}{linear feedback shift register}
\acro{LRA}{linear resonant actuator}
\acro{LSB}{least significant bit}
\acro{LUT}{look-up table}

\acro{MAC2}[MAC]{Multiply-Accumulate}
\acro{MIPS}{Microprocessor without Interlocked Pipeline Stages}
\acro{ML}{machine learning}
\acro{MMIO}{memory mapped \acl{IO}}
\acro{MUX}{multiplexer}
\acroplural{MUX}[MUXes]{multiplexers}
\acro{MSB}{most significant bit}

\acro{NASA}{National Aeronautics and Space Administration}
\acro{NDA}{non-disclosure agreement}
\acro{NMI}{normalized mutual information}
\acro{NSA}{National Security Agency}
\acro{NVM}{non-volatile memory}

\acro{OFB}{Output Feedback Mode}
\acro{OISC}{One Instruction Set Computer}
\acro{ORAM}{Oblivious Random Access Memory}
\acro{OS}{Operating System}
\acro{OSI}{Open Systems Interconnection}

\acro{PAR}{place-and-route}
\acro{PCB}{printed circuit board}
\acro{PC}{personal computer}
\acro{PDK}{process design kit}
\acro{PI}{proportional-integral}
\acro{PID}{proportional-integral-derivative}
\acro{PLB}{Programmable Logic Block}
\acro{PS}{passive serial}
\acro{PUF}{physical unclonable function}

\acro{RAM}{random-access memory}
\acro{RE}{reverse engineering}
\acro{RISC}{reduced instruction set computer}
\acro{RnD}[R\&D]{research and development}
\acro{RNG}{random number generator}
\acro{ROM}{read-only memory}
\acro{RTL}{register transfer level}

\acro{SAT}{Boolean satisfiability problem}
\acro{SCA}{side-channel analysis}
\acro{SCC}{strongly connected component}
\acro{SEM}{scanning electron microscope}
\acro{SE}{symbolic execution}
\acro{SHA}{Secure Hash Algorithm}
\acro{SMT}{satisfiability modulo theories}
\acro{SNR}{signal-to-noise ratio}
\acro{SOC}[SoC]{system-on-chip}
\acro{SPA}{simple power analysis}
\acro{SPI}{Serial Peripheral Interface}
\acro{SRAM}{static random access memory}
\acro{SRE}{software \acl{RE}}
\acro{STG}{state transition graph}

\acro{UART}{Universal Asynchronous Receiver Transmitter}
\acro{UHF}{ultra-high frequency}
\acro{USB}{Universal Serial Bus}

\acro{VHDL}{Very High Speed Integrated Circuit Hardware Description Language}
\end{acronym}

\ifdefined\algorithmautorefname
\renewcommand{\algorithmautorefname}{Algorithm}
\fi 
\ifdefined\definitionautorefname
\renewcommand{\definitionautorefname}{Definition}
\fi

\newcommand{\autorefapp}[1]{\hyperref[#1]{Appendix~\ref*{#1}}}

\newcommand{\HAL}{\texttt{HAL}\xspace}
\newcommand{\DANA}{\texttt{DANA}\xspace}

\newcommand{\etal}{et~al.\xspace}
\newcommand{\eg}{e.g.\xspace}
\newcommand{\ie}{i.e.\xspace}

\usepackage{pifont}
\definecolor{darkgreen}{RGB}{6, 156, 47}
\definecolor{darkred}{RGB}{240, 2, 2}
\newcommand{\cmark}{{\color{darkgreen}\ding{51}}}
\newcommand{\dmark}{{\color{orange}\ding{51}}}
\newcommand{\xmark}{{\color{darkred}\ding{55}}}

\usepackage{circledtext}
\circledtextset{boxfill=black}
\circledtextset{boxtype=o}
\circledtextset{charf=\bfseries\Large}
\circledtextset{charcolor=white}
\circledtextset{charshrink=0.6}
\circledtextset{resize=real}
\circledtextset{width=0.8em}

\begin{document}

\title[Stealing Maggie's Secrets---On the Challenges of IP Theft Through FPGA Reverse Engineering]{Stealing Maggie's Secrets---On the Challenges of IP Theft Through FPGA Reverse Engineering}

\author{Simon Klix}
\orcid{0000-0002-9369-2901}
\affiliation{%
  \institution{MPI-SP}
  \city{Bochum}
  \country{Germany}
}

\author{Nils Albartus}
\orcid{0000-0003-2449-1134}
\affiliation{%
  \institution{MPI-SP}
  \city{Bochum}
  \country{Germany}
}

\author{Julian Speith}
\orcid{0000-0002-8408-8518}
\affiliation{%
  \institution{MPI-SP}
  \city{Bochum}
  \country{Germany}
}

\author{Paul Staat}
\orcid{0000-0002-7539-4847}
\affiliation{%
  \institution{MPI-SP}
  \city{Bochum}
  \country{Germany}
}

\author{Alice Verstege}
\orcid{0009-0005-2807-0806}
\affiliation{%
  \institution{MPI-SP}
  \city{Bochum}
  \country{Germany}
}

\author{Annika Wilde}
\orcid{0000-0002-6092-7866}
\affiliation{%
  \institution{Ruhr University Bochum}
  \city{Bochum}
  \country{Germany}
}

\author{Daniel Lammers}
\orcid{0000-0002-9134-8568}
\affiliation{%
  \institution{Ruhr University Bochum}
  \city{Bochum}
  \country{Germany}
}

\author{Jörn Langheinrich}
\orcid{0000-0002-8583-5503}
\affiliation{%
  \institution{MPI-SP}
  \city{Bochum}
  \country{Germany}
}

\author{Christian Kison}
\orcid{0000-0002-5830-7692}
\affiliation{%
  \institution{Bundeskriminalamt}
  \city{Wiesbaden}
  \country{Germany}
}

\author{Sebastian Sester-Wehle}
\orcid{0000-0002-4938-2084}
\affiliation{%
  \institution{Bundeskriminalamt}
  \city{Wiesbaden}
  \country{Germany}
}

\author{Daniel Holcomb}
\orcid{0000-0002-2052-9820}
\affiliation{%
  \institution{UMass Amherst}
  \city{Amherst}
  \state{MA}
  \country{USA}
}

\author{Christof Paar}
\orcid{0000-0001-8681-2277}
\affiliation{%
  \institution{MPI-SP}
  \city{Bochum}
  \country{Germany}
}

\renewcommand{\shortauthors}{Klix, Albartus, Speith et al.}

\begin{abstract}
Intellectual Property (IP) theft is a cause of major financial and reputational damage, reportedly in the range of hundreds of billions of dollars annually in the U.S. alone.
Field Programmable Gate Arrays (FPGAs) are particularly exposed to IP theft, because their configuration file contains the IP in a proprietary format that can be mapped to a gate-level netlist with moderate effort.
Despite this threat, the scientific understanding of this issue lacks behind reality, thereby preventing an in-depth assessment of IP theft from FPGAs in academia. 
We address this discrepancy through a real-world case study on a Lattice iCE40 FPGA found inside iPhone~7.
Apple refers to this FPGA as \emph{Maggie}.
By reverse engineering the proprietary signal-processing algorithm implemented on Maggie, we generate novel insights into the actual efforts required to commit FPGA IP theft and the challenges an attacker faces on the way.
Informed by our case study, we then introduce generalized netlist reverse engineering techniques that drastically reduce the required manual effort and are applicable across a diverse spectrum of FPGA implementations and architectures.
We evaluate these techniques on six benchmarks that are representative of different FPGA applications and have been synthesized for Xilinx and Lattice FPGAs, as well as in an end-to-end white-box case study.
Finally, we provide a comprehensive open-source tool suite of netlist reverse engineering techniques to foster future research, enable the community to perform realistic threat assessments, and facilitate the evaluation of novel countermeasures.
\end{abstract}




\maketitle

\section{Introduction}
\label{maggie23::sec::introduction}
Many companies invest heavily into creating \ac{IP}, which often forms the backbone of their economic value and competitive advantage. 
For instance, Apple alone spent over \$26 billion on \ac{RnD} in 2022~\cite{companies_rd}.
Not surprisingly, valuable \ac{IP} can nowadays often be found within electronics and \acp{IC} in particular. 
It can take the form of proprietary algorithms for domains like \ac{DSP}, CPUs, or security functions.
As a flip side of this development, \ac{IP} theft, \eg, by competitors or hostile nation states, has become a major issue. 
In the U.S. alone, \ac{IP} theft causes hundreds of billions of dollars in damages every year, with the semiconductor market being of particular concern~\cite{li2020intellectual}.
In this paper, we look at an important aspect of this issue, namely assessing the security of \ac{IP} in \acp{FPGA}, which are, as we argue, particularly vulnerable against \ac{IP} theft.

In many highly specialized domains---such as spaceflight, military communications, network routers, or medical devices---\acp{FPGA} are a vital component, often incorporating proprietary \ac{IP}. 
\acp{FPGA} are \mbox{(re-)configurable} logic devices that are programmed using a bitstream with a proprietary file format.
This bitstream is generated by \ac{EDA} tools provided by the \ac{FPGA} vendors.
These tools map a high-level design, written in a \ac{HDL} such as Verilog, to a bitstream during synthesis.
At the end of this process, the implemented \ac{IP} is fully encapsulated in the bitstream file.
This file is stored externally to the \ac{FPGA} but resides on the target device. 

Improving our understanding of the associated challenges and required efforts is crucial to better assess the threat of, and defend against, real-world \ac{IP} theft from \ac{FPGA} implementations. 
In most cases, hardware \ac{IP} theft entails reverse engineering the design in order to recover the implemented algorithms and their instantiation.

Modern nanometer-scale \acp{IC} come with a natural defense against reverse engineering due to the high cost and labor-intensive nature of the netlist extraction process~\cite{DBLP:conf/aspdac/LippmannWUSEDGR19,DBLP:conf/ivsw/FyrbiakSKWERP17,DBLP:journals/jetc/QuadirCFASWCT16,DBLP:conf/dac/TorranceJ11,DBLP:conf/ches/TorranceJ09}---a financial barrier that deters all but the most resourceful entities, such as nation-states.
However, the bar for \ac{FPGA} reverse engineering is much lower.
To this end, an attacker would need to access the bitstream and map it to a gate-level netlist. 
This process is already well understood despite the proprietary nature of the bitstream format~\cite{DBLP:conf/aspdac/EnderSWWKP19,DBLP:conf/fpl/ZienerAT06,DBLP:conf/fpga/NoteR08,DBLP:conf/fpl/BenzSH12,DBLP:journals/mam/DingWZZ13,DBLP:conf/date/PhamHK17,debit,xray_github}. 
\ac{FPGA} vendors commonly provide cryptographic bitstream protections as a safeguard for valuable \ac{IP}.
Nonetheless, these measures have repeatedly been demonstrated to be susceptible to side-channel attacks~\cite{DBLP:conf/ccs/MoradiBKP11,DBLP:conf/ctrsa/MoradiKP12,DBLP:conf/cosade/0001S16,DBLP:journals/trets/SwierczynskiMOP15,DBLP:conf/ccs/TajikLSB17} or protocol flaws~\cite{DBLP:conf/uss/Ender0P20,DBLP:conf/fccm/EnderLMP22}.
Many low-cost \acp{FPGA} do not even offer such protections.
Even if these protective measures are available and resist such attacks, they must be \emph{proactively} enabled by the \ac{FPGA} developer.
Hence, we assume access to the plaintext bitstream, and thus the gate-level netlist, is generally achievable in practice.

In this situation, a reverse engineer is typically faced with a gate-level netlist that lacks any hierarchy, module boundaries, and word-level information such as data types, see \autoref{maggie23::subsec::challenges}.
Many netlist reverse engineering techniques have tried to address these challenges in recent years~\cite{DBLP:journals/jce/AzrielSAGMP21}, but they often deal only with isolated sub-problems such as register recovery, lack reference implementations, and use unrealistic and/or outdated benchmarks, see~\autoref{maggie23::subsec::related_work}.
In particular, an end-to-end analysis of a real-world \ac{FPGA} design remains largely unexplored in literature. 
As a consequence, the actual threat potential of \ac{FPGA} IP theft as well as the unique challenges it entails remain unknown.
Thus, the motivation of the paper at hand is to investigate the reverse engineering of a complex \ac{FPGA} netlist that is recovered from a real-world device in a black-box setting. 
The goal is to obtain and analyze the algorithmic description of the design. 
A better understanding of this issue can aid in the design of sound defenses and inform risk analysis of industry as well as governments.
\ac{FPGA} reverse engineering also paves the way for attacks other than \ac{IP} theft, \eg, retrieving hard-coded cryptographic keys or inserting hardware Trojans. 

\subsubsection*{Research Questions and Contributions}
For a realistic assessment of real-world \ac{FPGA} reverse engineering challenges in the context of \ac{IP} theft, we first set out to answer the following research question:

\vspace{0.1cm}
\begin{mdframed}
    \begin{questions}
        \item \label{maggie23::rq::1}
        What challenges and efforts are entailed with \ac{IP} theft through \ac{FPGA} reverse engineering in a black-box setting?
    \end{questions}
\end{mdframed}
\vspace{0.1cm}

To answer \ref{maggie23::rq::1}, we demonstrate the steps and efforts required to extract \ac{IP} from an \ac{FPGA} implementation through a case study on a Lattice iCE40 \ac{FPGA} found within the iPhone~7.
To this end, we combine existing techniques and propose new approaches to recover an algorithmic description of the target \ac{FPGA} implementation, effectively extracting the \ac{IP} controlling the iPhone's \emph{Taptic Engine}, see \autoref{maggie23::sec::case_study}.
In addition to evaluating existing techniques and proposing novel approaches to \ac{FPGA} reverse engineering, we provide unique insights that foster developing better defenses against \ac{IP} theft.
While the iPhone~7 was released in 2016, Lattice iCE40 \acp{FPGA} are still used in many other devices such as Apple Vision Pro, Apple TV 4K,  HTC Vive Pro, Samsung Galaxy S5, and Pebble Time.
As the bitstreams of all Lattice iCE40 \acp{FPGA} follow the same structure, our tooling could be directly applied to the bitstreams of these other devices to obtain their gate-level netlists.
Subsequently, the netlist reverse engineering techniques presented in this paper could be used to analyze the extracted netlist right away.

Please note that Apple refers to the \ac{FPGA} inside iPhone~7 as \emph{Maggie} in the firmware, and we will also use this name throughout our paper.
Based on our findings, a second research question arises: 

\vspace{0.1cm}
\begin{mdframed}
    \begin{questions}
        \setcounter{questionsi}{1}
        \item \label{maggie23::rq::2} 
        To what extent can \ac{FPGA} reverse engineering be generalized across architectures and implementations?
    \end{questions}
\end{mdframed}
\vspace{0.1cm}

In response to \ref{maggie23::rq::2}, we generalize the methods developed for our case study. 
This yields automated methods that are applicable across \ac{FPGA} architectures and implementations. 
In particular, we target Xilinx 7-Series and Lattice iCE40 \ac{FPGA} families, see \autoref{maggie23::sec::generalization}.
We evaluate the presented techniques in isolation on a diverse set of benchmark netlists and discuss differences in effectiveness.
Additionally, we conduct an end-to-end white-box case study to demonstrate the effectiveness of our workflow as a whole.
In line with the efforts of the security community to provide open-source implementations, thereby removing financial barriers and enabling third-party verification, we share our benchmarks and implementations as plugins to \HAL~\cite{hal_github}.
Thereby, we hope to encourage more insight into \ac{FPGA} reverse engineering in follow-up work by academia and industry, and foster the development of countermeasures, see \autoref{maggie23::sec::discussion}.

\subsubsection*{Responsible Disclosure}
Despite not revealing any security vulnerabilities, we informed Apple and Lattice Semiconductor about our findings before publication.
In agreement with the vendors, we will not publish the gate-level netlist recovered from Maggie's bitstream or our bitstream conversion tooling for Lattice iCE40 \acp{FPGA}.

\section{Attacker Model \& Challenges}
\label{maggie23::sec::attacker}

\subsection{Attacker Model}
\label{maggie23::subsec::attacker}
Our attacker model revolves around \ac{IP} theft, which is a common threat considered in hardware security research~\cite{forte2017hardware,DBLP:conf/host/ShamsiLMZPJ17}.
An attacker obtains a device featuring an \ac{FPGA} and aims to recover the contained \ac{IP} by extracting an algorithmic model of the design implemented on the \ac{FPGA} by means of reverse engineering.
We assume the attacker has gained access to the bitstream configuring the \ac{FPGA} by either extracting it from the device or from firmware.
Given that the attacker possesses the device, they can execute the \ac{FPGA} implementation and modify the hardware as needed.

As is typical in reverse engineering, the attacker may have already gathered public information on the target device, such as documentation or patents.
This leaves them with knowledge of the system's connectivity and interaction with the \ac{FPGA}, along with a general understanding of its purpose, but lacking detailed knowledge of the implementation and the intricacies of the algorithm.
Ultimately, the attacker seeks to reconstruct a detailed algorithmic representation that allows him to extract, understand, and modify the \ac{IP} contained within it.
Thus, our attacker model explicitly goes beyond simply copying an \ac{FPGA} bitstream or netlist. 

\subsection{Netlist Reverse Engineering Terminology}
\label{maggie23::subsec::terminology}
A gate-level netlist is a digital circuit representation comprising combinational and sequential \emph{gates} or standard cells as well as their interconnections, also known as \emph{nets}.
In reverse engineering, we frequently encounter flattened netlists lacking hierarchy or module boundaries.
Furthermore, we differentiate between \textit{data path} and \textit{control path}. 
The control path steers the data path in that it decides which operations are being performed on the data passing through by enabling or disabling certain parts of the circuit.

Reverse engineering gate-level netlists involves \textit{static} and \textit{dynamic} techniques.
Static methods analyze the netlist graph and can take either \textit{structural} or \textit{functional} information into account.
Dynamic techniques leverage data gathered during execution to reverse engineer functionality.

\subsection{Challenges of Netlist Reverse Engineering}
\label{maggie23::subsec::challenges}
Committing \ac{IP} theft on an \ac{FPGA} requires analyzing the gate-level netlist extracted from the bitstream, posing unique challenges. 
These challenges arise from the need to reconstruct information removed during synthesis.
The challenges we faced include:

\begin{challenges}
    \item \label{maggie23::ch::1} \textbf{Loss of Hierarchy.}
    \ac{HDL} designs feature modules that encapsulate functionality and introduce hierarchy.
    This information is removed during synthesis.
    The resulting netlist lacks any module boundaries or hierarchy.
    
    \item \label{maggie23::ch::2} \textbf{Loss of Data Types.}
    An \ac{HDL} description makes use of multi-bit data types like integers on which, \eg, arithmetic operations are performed.
    In contrast, gate-level netlists only contain gates and nets operating on bit level.
    Not only is the information of which gates belong to the same word-level operation lost during synthesis, but also the order of bits within word-level structures.
    Consequently, high-level data types no longer exist in the netlist which hampers analysis relying on, \eg, integer values.

    \item \label{maggie23::ch::3} \textbf{Synthesizer Optimizations.}
    Synthesizer optimizations can result in similar operations on \ac{HDL} level looking vastly different in a gate-level netlist.
    Logic may be merged across module boundaries and gates, or entire sub-circuits may be duplicated to facilitate a high fan-out.
    Such optimizations always depend on the available standard cells, timing and area constraints, routing requirements, and varying optimization strategies.

    \item \label{maggie23::ch::4} \textbf{Missing Control Separation.} 
    The control path is often implemented using \acp{FSM}.
    In an \ac{HDL} design, \acp{FSM} can be distinguished from one another and the data path.
    In netlists, \acp{FSM} are often merged with one another or even the data path such that boundaries can no longer be determined.
    This causes a state explosion when analyzing affected \ac{FSM} state graphs, sometimes raising complexity beyond what is feasible.
    
    \item \label{maggie23::ch::5} \textbf{Data Dependency.} 
    In our case study, the behavior of the control path---and in extension the data path---depended on external data and instructions fed to the \ac{FPGA} as input.
    Hence, analyzing such a netlist without input data is challenging, especially if the control path cannot be properly dissected. 

    \item \label{maggie23::ch::6} \textbf{Dynamic Behavior.} 
    Sequential gates add another dimension to netlist reverse engineering in that they require the reverse engineer to consider past states for the analysis of current behavior.
    This blows up complexity due to the vast number of possible netlist states to consider.

    \item \label{maggie23::ch::7} \textbf{Semantic Analysis.}
    Extracting a word-level algorithmic representation leaves a reverse engineer with the task of assigning meaning and symbols to different values of the computation. 
    To gain a high-level understanding, including the underlying rationale and specific design decisions, the functionality needs to be dissected rigorously from the system level down to individual building blocks, calling for domain-specific expertise.
\end{challenges}

\section{Case Study on iPhone 7}
\label{maggie23::sec::case_study}
To answer \ref{maggie23::rq::1} and investigate the effort required to commit \ac{IP} theft from a real-world \ac{FPGA} implementation, we conducted a case study on an \ac{FPGA} found in iPhone~7.

\subsubsection*{Target Device}
The iPhone~7 and iPhone~7~Plus were released in September 2016.
The smartphones feature a novel capacitive home button with haptic feedback being generated by the \textit{Taptic Engine} first introduced in Apple Watch and iPhone~6s.
Notably, both iPhone~7 and iPhone~7~Plus contain a Lattice iCE5LP4K-SWG36 \ac{FPGA}.
Apple refers to this chip as \textit{Maggie} throughout firmware and leaked \ac{PCB} schematics.
There are many speculations revolving around the purpose of Maggie~\cite{Martellaro2016,Tilley2016,ifixit2016}.

\begin{figure}[!htb]
  \centering
  \includegraphics[width=0.95\linewidth]{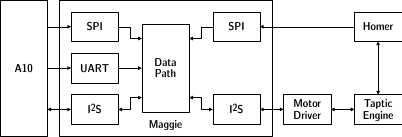}
  \caption{Maggie as part of the Taptic Engine controller.}
  \label{maggie23::fig::maggie_system_overview}
  \Description[<short description>]{<long description>}
\end{figure}

\subsubsection*{Setting}
\begin{figure}[!htb]
  \centering
  \includegraphics[width=0.85\linewidth]{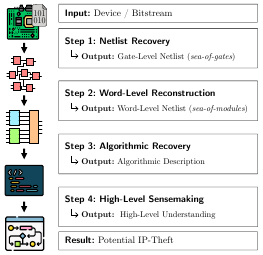}
  \caption{Overview of our case study on iPhone~7.}
  \label{maggie23::fig::iphone_case_study_overview}
\end{figure}

From analyzing leaked \ac{PCB} schematics, the iPhone firmware, and patents granted to Apple~\cite{Hajati2021,Hajati2018}, we concluded that Maggie controls the \textit{Taptic Engine}, and determined its \acs{IO} connectivity and communication interfaces, see \autoref{maggie23::fig::maggie_system_overview}.
We further discovered that the \ac{FPGA} itself is programmed on start-up and is not used as a re-programmable accelerator; it always implements the same control logic of the Taptic Engine, and the design does not differ between applications.
The \ac{FPGA} is controlled by the A10 \ac{SOC} and triggered by the OS to generate haptic feedback, \eg, to simulate home button presses.
However, this knowledge alone does not enable effective \ac{IP} theft, as we still lack implementation details on the instantiated signal processing algorithms and their parameters.
Assuming the role of the attacker, we therefore need to investigate the circuit implemented on Maggie to fully recover Apple's Taptic Engine \ac{IP}.
To this end, we split the iPhone \ac{FPGA} reverse engineering process into four steps, see \autoref{maggie23::fig::iphone_case_study_overview}.

\subsection{Step 1: Netlist Recovery}
\label{maggie23::subsec::netlist_recovery}
At first, we were faced with the iPhone~7 containing the bitstream featuring the target \ac{IP} and our first goal was to extract the gate-level netlist for further analysis.

\subsubsection*{Bitstream Acquisition}
\acs{SRAM}-based \acp{FPGA} such as Lattice iCE40 devices store the bitstream externally and are configured during startup.
Hence, an attacker can read the bitstream from external memory, intercept it during transmission, or recover it from device firmware.
The Maggie bitstream is shipped with the openly available iPhone firmware, hence we extracted it from the file system. 
The original bitstream was published in September 2016 and was updated with iOS~10.2 in December 2016.
As it has remained unchanged ever since, our analysis deals with this latest version.
As Lattice iCE40 \acp{FPGA} do not offer any bitstream protections such as encryption, we were directly faced with then plaintext bitstream.
This lack of security features can occasionally be observed in low-cost \acp{FPGA}, leaving the \ac{IP} they implement mostly unprotected.
Please refer to \autoref{maggie23::app::bitstream_enc} for a list of \acp{FPGA} that do (not) support bitstream encryption.

\subsubsection*{Bitstream Format Reverse Engineering}
The bitstream format is typically a secret well-kept by \ac{FPGA} vendors, although everyone can generate bitstreams using respective \ac{EDA} tools.
Project IceStorm~\cite{icestorm_github} entails a comprehensive documentation database of the Lattice iCE40 bitstream format generated by fuzzing the Lattice iCEcube2 \ac{EDA} tool.
Project IceStorm is part of F4PGA~\cite{F4PGA2022}, an open-source \ac{EDA} tool-flow developed as an alternative to proprietary vendor software. 
As part of our efforts, we extended Project IceStorm by increasing its fuzzing coverage and adding support for the pinout of Maggie's SWG36 chip package.

\subsubsection*{Bitstream Conversion}
Given a database detailing the mapping from bits to netlist elements, an arbitrary bitstream can be converted back to a gate-level netlist in full automation.
In light of our case study, we developed a custom bitstream conversion tool that translates any Lattice iCE40 bitstream into a gate-level netlist.
Our tool depends on the Project IceStorm database to be complete and correct.
For Maggie, we observed 36 nets with missing sources or destinations, hinting at an incomplete bitstream database.
Later on, we manually repaired these nets using information on surrounding circuitry revealed once our analysis progressed.
The extracted netlist contains 5046 gates comprising 3241 \acp{LUT}, 422 carry gates, 1342 \acp{FF}, 18 \acp{BRAM}, and 4 \ac{DSP} units.

\subsection{Step 2: Word-Level Reconstruction}
\label{maggie23::subsec::word_level_rec}
After extracting the gate-level netlist, we aimed to recover word-level structures from the unstructured sea-of-gates using the netlist reverse engineering framework \HAL~\cite{hal_github}.
To this end, we were confronted with the challenges outlined in \autoref{maggie23::subsec::challenges}.

\subsubsection*{Netlist Pre-Processing}
We initially pre-processed the netlist to eliminate synthesis byproducts hindering analysis and enhance readability for human reverse engineers.

\begin{figure}[htb!]
     \centering
     \begin{subfigure}[t]{0.48\linewidth}
         \centering
         \includegraphics[width=0.85\linewidth]{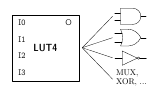}
         \caption{Replace \acp{LUT} with primitive combinational gates.}
         \label{maggie23::fig::lut_decomp_simple}
     \end{subfigure}
     \hfill
     \begin{subfigure}[t]{0.48\linewidth}
         \centering
         \includegraphics[width=0.85\linewidth]{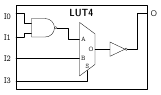}
         \caption{Decompose \acp{LUT} into a \acs{MUX} and primitive gates.}
         \label{maggie23::fig::lut_decomp_nested}
     \end{subfigure}
    \caption{Netlist Preparation Overview.}
    \label{maggie23::fig::lut_decomp}
\end{figure}

For more efficient manual analysis and considering that some \acsp{LUT} merely implement basic functions like AND or XOR, we substituted such \acsp{LUT} with primitive gates, see \autoref{maggie23::fig::lut_decomp_simple}.
We also removed buffer \acsp{LUT} as they add no functionality but sometimes hinder structural analysis.
Furthermore, we removed gates and sub-circuits that compute the same function on identical inputs.
The synthesizer introduces such duplicates for optimization purposes, yet they can obfuscate shared control signals or data inputs.

\acsp{MUX} in particular provide structure to the data path.
On \acp{FPGA}, however, their functionality is implemented as part of more complex Boolean functions within \acp{LUT}.
Hence, decomposing complex \acsp{LUT} into small building blocks such as \acsp{MUX} can aid manual and automated reverse engineering, \eg, by supporting structural analysis.
To recover nested \acsp{MUX} and better understand the data path (see \autoref{maggie23::fig::p2_muxes}), we analyzed each \ac{LUT}'s Boolean function and searched for inputs that behave like a select.
When discovering a \acs{MUX} in a \acs{LUT}, we replaced the \acs{LUT} with a \acs{MUX} and reconstructed the surrounding logic from primitive gates. 
In total, we replaced 549 \acsp{LUT} with primitive gates, removed 594 buffers, extracted 1619 \acsp{MUX} and 1161 primitive gates, removed 353 duplicate gates, and removed 51 additional gates using other simplifications.

\subsubsection*{Control Logic}
Analyzing Maggie's control path was required for comprehending how the circuit operates.
Additionally, certain techniques operating on the data path require knowledge of which gates and nets belong to the control path.
Unfortunately, the absence of \ac{FPGA}-compatible reference implementations~\cite{DBLP:journals/jhss/MeadeSLDZJ18,DBLP:conf/aspdac/MeadeZJ16,DBLP:conf/dac/GeistMZJ20}, the absence of public implementations altogether~\cite{DBLP:conf/host/BrunnerBS19, DBLP:journals/tches/FyrbiakWDABTP18,DBLP:conf/iscas/ShiTGR10,DBLP:journals/todaes/BrunnerHBS22}, or general assumptions on \acp{SCC} that do not always hold for our case study~\cite{DBLP:conf/iscas/ShiTGR10,DBLP:journals/tches/FyrbiakWDABTP18}, prevented us from applying existing techniques from literature.
In particular, the \texttt{NETA} toolset~\cite{neta_github} is only available as binaries that cannot parse \ac{FPGA} netlists.
Moreover, a common assumption is that the state registers and transition logic of an \acp{FSM} form an \ac{SCC}~\cite{DBLP:conf/iscas/ShiTGR10,DBLP:journals/tches/FyrbiakWDABTP18}.
For Maggie, we found 55 \acp{SCC} with all but seven comprising three gates or less.
Vast parts of the netlist form a single \ac{SCC} of more than 5000 gates.
Therefore, \acp{SCC} are not useful in our case study.

Given the intricacies of this issue, we could not find a workable solution within the scope of this work and hence decided to manually investigate the control path of Maggie instead.
Here, we faced significant challenges due to the \acsp{FSM} being interwoven to an extent that often prevented even manual separation, as discussed in \cite{DBLP:journals/todaes/BrunnerHBS22}.
While \ac{FSM} separation was not always feasible, we still classified \acsp{FF} as part of the control path by checking whether they feed into control pins such as \acs{MUX} select or \acs{FF} enable.
After assigning a \ac{FF} to the control path, we examined its predecessors and successors.
These structural characteristics are merely clues that a gate might belong to the control path.
They often required manual inspection and human intuition, hence preventing full automation.

\subsubsection*{Communication Interfaces}
Investigating Maggie's \acs{SPI}, \acs{UART}, and \acs{I2S} interface implementations (see \autoref{maggie23::fig::maggie_system_overview}) supports word-level structure recovery as these interfaces often comprise the first or last register of a data path.
Furthermore, we must reverse engineer these interfaces to understand the parallel and serial conversion of incoming and outgoing data, which is a precondition for recovering bit orders and data types.
Through manual reverse engineering, we identified 615 gates belonging to five interfaces.

\subsubsection*{Arithmetic Structures}
\begin{figure}[!htb]
    \centering
    \includegraphics[width=0.9\linewidth]{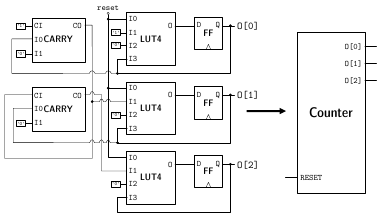}
    \caption{Structure of a $3$-bit counter with reset constructed from a carry chain on a Lattice iCE40 \ac{FPGA}.}
    \label{maggie23::fig::p2_counter_structure}
\end{figure}

In the iPhone netlist, we observed the data path comprising mostly combinational logic implemented as \acsp{LUT} and carry gates as well as sequential logic implemented using \acsp{FF}.
Some data-path structures are implemented in a consistent, recurring way, \eg, adders, counters, and selected comparators.
Lattice iCE40 \acp{FPGA} provide carry gates that are commonly connected to form carry chains, often used in combination with \acp{LUT} to implement arithmetic operations, see \autoref{maggie23::fig::p2_counter_structure}.
A reverse engineer can leverage these recurring patterns to identify word-level structures like adders or counters in a gate-level netlist.

To recover word-level arithmetic operations, we first located carry chains in the netlist by scanning for connected carry gates. 
We then assembled arithmetic structure candidates by adding surrounding logic to the carry chains.
Since a structure may be influenced by control inputs, we constructed structural candidate variants to account for different control behaviors.
Operand membership and the order of input and output signals were derived from the structure and connectivity of a candidate's gates.
For example, the order of inputs was reconstructed by identifying the carry gate of the chain to which a signal is connected.
Finally, we checked each candidate against predefined \ac{SMT} models, \eg, $A \pm B$ for adders or $A \pm n$ for counters incrementing by a constant $n$.

\subsubsection*{Sequential Data Path}
The synthesizer realizes word-level registers from the \ac{HDL} design as individual \acsp{FF} that lack any indication of which register they belong to.
To this end, \DANA~\cite{DBLP:journals/tches/AlbartusHTAP20} can be used to recover these word-level registers from a gate-level netlist using structural metrics such as shared predecessors and successors as well as common control signals. 
However, as \DANA is optimized for data-path analysis, control \acp{FF} may not be recognized as such and instead be wrongfully merged into data-path registers.
This impairs the detection of other registers due to \DANA's iterative nature.
Hence, we extended \DANA to include \textit{known registers} identified by other analysis techniques for finding, \eg, communication interfaces, \acp{FSM} states, and counters.
\DANA does not alter these registers but utilizes them to identify registers further up or down the data path.

\subsubsection*{Combinational Data Path}
\begin{figure}[!htb]
  \centering
  \includegraphics[width=0.8\linewidth]{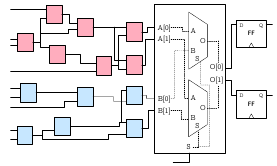}
  \caption{The recovery of word-level \acsp{MUX} allows for the analysis of separate independent data paths.}
  \label{maggie23::fig::p2_muxes}
\end{figure}

In the data path, \acsp{MUX} can be used to switch between word-level data sources, such as global inputs, memory, or arithmetic structures.
We implemented a method to group \acsp{MUX} that have been decomposed from \acp{LUT} as part of our pre-processing, see \autoref{maggie23::fig::lut_decomp}.
A \acs{MUX} selects an input to be forwarded to the output based on a select input \texttt{S}.
Multiple \acsp{MUX} sharing a common select typically form a word-level \acs{MUX} structure.
Detecting such structures allows for the separation of word-level data paths, see \autoref{maggie23::fig::p2_muxes}.
However, grouping by select can yield oversized structures as the same select signal may influence two or more distinct data paths.
Hence, we split these groups according to preceding or successive word-level structures.
For example, if a 32-bit \acs{MUX} was succeeded by two 16-bit registers, the \acs{MUX} was split into two 16-bit \acsp{MUX}.
As other preceding or successive \acsp{MUX} themselves were considered for splitting, the process was repeated iteratively until no further changes could be observed.

\subsubsection*{Bit Order}
While the order of signals (\ie, the \textit{bit order}) belonging to a word is inherently known in the \ac{HDL} design, it is removed during synthesis.
Hence, for a reverse engineer it is not clear which bits of, \eg, a register are the \acs{LSB} or \acs{MSB}.
Recovering the bit order of word-level structures aids manual investigation of the design and is vital for the reconstruction of data types used during simulation, see \autoref{maggie23::subsec::algorithmic_rec}.
For adders, counters, \acsp{BRAM}, and \acsp{DSP}, the bit order can be inferred from their topology and function.
Implementations of other structures, \eg, word-level registers and \acsp{MUX}, do not exhibit an obvious bit order.

\begin{figure}[!htb]
  \centering
  \includegraphics[width=0.85\linewidth]{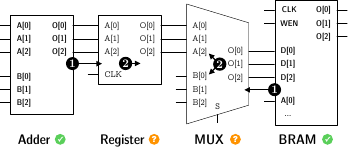}
  \caption{
      The bit orders of the adder and the \acs{BRAM} are known, but not those of the register and the \acs{MUX}. 
      In iteration~\circledtext{1}, the order of adder pin group \texttt{O} is propagated to register pin group \texttt{A}; the order of \acs{BRAM} pin group \texttt{D} is propagated to \acs{MUX} pin group \texttt{O}.
      In iteration~\circledtext{2}, both orders are propagated within the register and the \acs{MUX} to cover their outputs and inputs. 
      Finally, the algorithm terminates as all bit orders have been annotated.
  }
  \label{maggie23::fig::bitorder_propagation}
\end{figure}

We propagated the bit order from structures with known bit order along the data path to structures with unknown bit order.
For every unordered pin group, our algorithm ($i$)~derives bit-order candidates from predecessors and successors and ($ii$)~tries to establish consensus between them.
Candidates with inherent conflict are discarded, \eg, when different pins of a pin group fetched the same index from the same origin.
We omit control logic during analysis, as it does not feature a natural bit order and degrades results.

\subsubsection*{Outcome}
\autoref{maggie23::tab::recovered_structures} depicts the results for our automated techniques.
In addition, we manually grouped 688 gates into data-path modules that did not feature \acsp{MUX} and identified 631 gates belonging to control logic.
In total, we assigned 4570 out of 5496 gates (83\%) to structures using manual and automated techniques.
Furthermore, we assigned a bit order to 605 (71\%) out of 843 pin groups from word-level structures that were not classified as control logic.
While we could not verify against a ground truth as of the black-box nature of our case study, these results enabled correctly extracting the implemented algorithm, thereby underlining their value.

\begin{table}[!htb]
    \centering
    \begin{tabularx}{\linewidth}{Xrrc}
         \toprule
         \textbf{Structure}                    & \textbf{\#Recovered} & \textbf{\#Gates} & \textbf{Automated?}\\
         \midrule
         Arithmetic                            &           35 &           692 & \cmark \\
         Register                              &          118 &          1342 & \cmark \\
         Multiplexers                          &          160 &          1217 & \cmark \\
         \bottomrule
    \end{tabularx}
    \caption{Results of our word-level reconstruction.}
    \label{maggie23::tab::recovered_structures}
\end{table}

\subsection{Step 3: Algorithmic Recovery}
\label{maggie23::subsec::algorithmic_rec}
Having previously recovered word-level structures, we extracted an algorithmic description of what the iPhone netlist implements.
This entails further analyzing the control path, as we could not draw boundaries between individual control structures so far.

\subsubsection*{Virtual Probing}
\label{maggie23::subsec::algorithmic_rec::virtual_probing}
Instead of further attempting to unravel the semantics of the interwoven control logic, we decided to use dynamic reverse engineering techniques to observe the actual behavior of the circuit at runtime.
However, such dynamic analysis often requires expensive equipment and yields only incomplete insights into the execution state~\cite{DBLP:conf/host/NedospasovSSO12,DBLP:conf/ccs/TajikLSB17}.
Hence, we leveraged a more workable approach: netlist simulation using \acs{IO} signals captured during operation.
We further used this approach to ($i$) observe the flow of real-world data through the netlist, assisting in algorithmic recovery, and ($ii$) to verify correctness of the recovered netlist and the later extracted algorithm implemented on Maggie by comparing simulation results against recorded outputs.

\begin{figure}[!htb]
    \centering

    \includegraphics[width=0.75\linewidth]{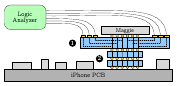}
    \caption{
        A breakout board is placed between Maggie and the iPhone~7 \ac{PCB} to record Maggie's \acs{IO}.
        The breakout board comprises two parts: \circledtext{1}~a wider \ac{PCB} passing the signals to Maggie and to a logic analyzer and \circledtext{2}~two small \ac{PCB} cubes vertically routing the signals
    }
    \label{maggie23::fig::pcb_probing}
\end{figure}

We captured \acs{IO} signals by placing a custom breakout board between Maggie and the iPhone \ac{PCB}, see \autoref{maggie23::fig::pcb_probing}.
By feeding the captured \acs{IO} signals to a simulator, we could then analyze the state of the circuit at any point in time.
This reduced complexity of the circuit's theoretical state space by focusing only on the states that are actually reached.
To this end, we extended the netlist analysis framework \HAL~\cite{hal_github} with simulation capabilities, see~\autoref{maggie23::app::hal_simulation}.
This approach allowed for, \eg, analysis of the data flow over time and correct traversal of states for complex \acp{FSM} that otherwise exhausted functional analysis.
Given the previously reconstructed word-level structures and their bit orders, we could now follow the flow of multi-bit data through the design and observe which operations were executed at what point in time.
This completely alleviated us from the need to statically analyze the control path.

\subsubsection*{DSP Analysis}
\begin{figure}[!htb]
     \centering
     \includegraphics[width=0.8\linewidth]{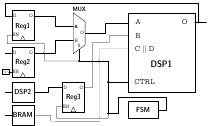}
     \caption{
        Simplified \acs{DSP} circuit. 
    }
    \label{maggie23::fig::dsp_circuit}
\end{figure}

By tracing inputs to Maggie through the netlist in simulation, we found that almost all data-path operations are performed by four connected \acsp{DSP}.
On Lattice iCE40 \acp{FPGA}, \acsp{DSP} feature four 16-bit data inputs and a 32-bit output~\cite{lattice_ice_technology_library}.
On Maggie, all \acsp{DSP} are configured as \ac{MAC2} units performing 16-bit multiplication and 32-bit accumulation.
Most \acs{DSP} inputs are preceded by \acp{LUT} implementing \acsp{MUX} to allow for dynamically changing data sources, see \autoref{maggie23::fig::dsp_circuit}.
Each \acs{DSP} and its \acsp{MUX} are controlled by an \ac{FSM}.
Looking at the simulation, we found that every \acs{DSP} computes a recurring sequence of operations depending on dynamically changing control and data inputs.

For each step of these sequences, we leveraged simulation to (i)~analyze control inputs and the internal \acs{DSP} state to determine the executed arithmetic operation and (ii)~identify the data sources which the operands of this operation came from.
The operands may have passed through \acsp{MUX} and can originate from communication interfaces, registers, \acsp{BRAM}, or other \acsp{DSP}.
We also encountered feedback loops when a \acs{DSP} output is stored in a register that is again applied as an input to the same \acs{DSP} later on, see \texttt{Reg1} of \autoref{maggie23::fig::dsp_circuit}.
Please note that we ignored the actual operand values, but rather traced their data sources to reconstruct symbolic equations describing the executed operations.

For this, we started at the inputs of the respective \acs{DSP}.
Whenever encountering a combinational structure such as a word-level \acs{MUX}, we looked at the current value of its select signal in simulation to determine the selected data path.
Sequential components add another dimension in that they require knowledge of past circuit states, which blows up complexity of static analysis.
However, we could leverage information from simulation to avoid investigating all possible circuit states.
For registers, we determined the simulation cycle in which they were last written and continued traversal from there.
We proceeded with this manual approach until we reached another \acs{DSP}, a memory, or a communication interface.
By applying this to all four \acsp{DSP}, we generated equations describing Maggie's data path.
Using these equations, we crafted a Python script that replicates the algorithm implemented by the data path and verified correctness by checking against recorded outputs.

\subsection{Step 4: High-Level Sense-Making}
\label{maggie23::subsec::high_level}
Based on the recovered algorithm, we proceeded with a semantic analysis that is essential for meaningful interaction with Apple's \ac{IP}.
Generating a Python script implementing the algorithm proved crucial since comprehending the algorithm required domain knowledge that was beyond our \ac{FPGA} reverse engineering team.
This way, we could share the Python script containing the reconstructed algorithm with a signal processing expert without requiring them to have any knowledge of the reverse engineering process.

\subsubsection*{Big Picture}
\begin{figure}[!ht]
  \centering
  \includegraphics[width=0.9\linewidth]{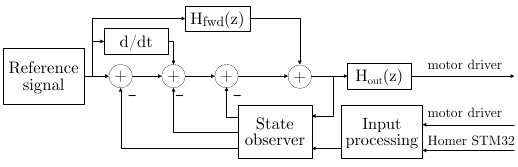}
  \caption{Overview of Maggie's signal processing.}
  \label{maggie23::fig::maggie_dsp_overview}
\end{figure}

Maggie is a central component of the iPhone~7 haptics subsystem to which Apple refers as \textit{Taptic Engine}.
We found that it controls the excitation of the~\ac{LRA} to produce a vibration which the user perceives as a tactile stimulus. 
For instance, the Taptic Engine creates the sensation of a physical button push for the non-mechanical home button of the iPhone~7. 
To provide such refined tactile feedback, a closed-loop control system is utilized, which makes the~\ac{LRA} accurately follow a desired movement.
Here, Maggie implements the closed-loop motion controller, see~\autoref{maggie23::fig::maggie_dsp_overview} for a simplified block diagram, which conditions the drive signals applied to the actuator (via the motor driver) based on sensor feedback from the actuator itself.

\subsubsection*{Signal Processing}
To generate the \ac{LRA}~control signals, Maggie operates on two feedback signals from the~\ac{LRA}, a reference signal previously loaded into \acs{BRAM}, and values of intermediate signals at previous sampling instants.
From the algorithmic representation, we identified several classical biquad filter stages implemented in a direct form~I~structure~\cite{oppenheim2014discrete}. 
Upon closer inspection, we found many typical \acs{DSP} building blocks, \eg, sub-sampling and sample-and-hold blocks for sample rate conversion, bit shifts and truncations for fixed-point arithmetic, saturation stages to prevent overflows, and filter implementations. 
Furthermore, we recognized subtle details such as the use of \textit{fraction saving}~\cite{Yates2008}. 
We could even identify the individual building blocks along with their value-exact parametrization, including filter and control-loop coefficients and initialization values---fully characterizing the entire signal processing chain.
Putting it all together, we found three main processing stages:
$(i)$~input signal conditioning and application of calibration, 
$(ii)$~a state observer to track the actuator dynamics, and 
$(iii)$~the closed-loop controller to minimize errors between the actual and the desired~\ac{LRA} movement. 
For additional details on the recovered \ac{DSP} operations, see \autoref{maggie23::app::dsp_details}.

\section{Deriving Generalized Techniques}
\label{maggie23::sec::generalization}
Our case study demonstrated that a semi-automated approach for \ac{FPGA} reverse engineering in the context of \ac{IP} theft entails considerable effort.
In line with \ref{maggie23::rq::2}, we now investigate whether (and to what extent) the learnings from our case study can be generalized across different \ac{FPGA} architectures.
Upon publication, we will provide open-source implementations of our generalized tools.

We evaluated the techniques discussed in this section on six benchmark designs synthesized for Xilinx 7-series and Lattice iCE40 \acp{FPGA}, see \autoref{maggie23::tab::eval} for evaluation results and \autoref{maggie23::app::benchmarks} for details on the benchmarks.
Furthermore, we conducted an end-to-end white-box case study on one signal processing design implemented for Xilinx 7-series \acsp{FPGA} to showcase the effectiveness of our generalized techniques in concert.

\subsection{Netlist Pre-Processing}
So far, our pre-processing steps were limited to the 4-input \acp{LUT} found on Lattice iCE40 \acp{FPGA}.
While some techniques such as the detection of duplicate gates could easily be extended to cover arbitrary combinational gates, this is not true for our \ac{LUT} decomposition.
Here, our approach works reasonably well for 4-input \acp{LUT} but does not scale to six inputs.
Hence, for generalization, we re-synthesize the combinational subgraphs between sequential gates using the open-source \ac{EDA} tool Yosys~\cite{DBLP:conf/fccm/ShahHWBGM19} instead of analyzing each \acs{LUT}'s Boolean function.
To this end, we constrain the synthesizer to a custom gate library comprising only the desired combinational gates such as \texttt{INV}, \texttt{BUF}, \texttt{AND}, \texttt{OR}, \texttt{XOR}, \texttt{XNOR}, and \texttt{MUX} of various sizes.
Compared to decomposing the combinational parts of the netlist into a pure \ac{AOI} graph, this approach allows us to delegate \ac{MUX} detection to the synthesizer by incentivizing the selection of \ac{MUX} gates over alternatives by artificially reducing their size and optimizing for area.
This way, we can even identify \acsp{MUX} within complex \ac{LUT} configurations.
Another benefit of using only primitive gates is that it improves the efficiency of structural analysis. 
Allowing more complex combinational gates during re-synthesis can result in the synthesizer merging interconnected logic into fewer gates, which obscures module boundaries and hampers structural analysis.
By nature, this re-synthesis approach is independent of the underlying \ac{FPGA} architecture and target implementation.

\subsubsection*{Evaluation}
We verified the correctness of our pre-processing using SMT solving.
However, pre-processing is a best-effort approach and lacks a ground truth to compare against.
It is essential for subsequent techniques such as word-level \acs{MUX} detection and arithmetic structure identification.
The success of these techniques serves as a proxy to gauge the effectiveness of our pre-processing approach.

\subsection{Arithmetic Structures}
\label{maggie23::subsec::arithmetic}

\begin{figure*}[!ht]
  \centering
  \includegraphics[width=1\linewidth]{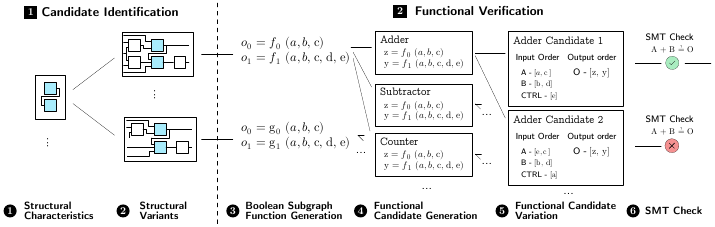}
  \caption{Module Identification and Classification Overview.}
  \label{maggie23::fig::module_identification}
\end{figure*}

In our case study, we identified arithmetic operations by building candidate sets of gates and verifying their function using an \ac{SMT} solver.
However, utilizing an \ac{SMT} solver to compare against functional models from a library of known operations requires knowledge of the order of input and output signals as well as their operand membership.
Furthermore, identifying any existing control signals presents additional challenges.
When analyzing Maggie, we inferred operands of arithmetic operations from structural properties such as the pins of carry gates, \eg, operand $A$ of an addition $A+B$ would often be connected to input \texttt{I0} of a Lattice iCE40 \texttt{CARRY} gate.
The bit order of arithmetic operations was deduced from the structure of the underlying carry chains.
This approach is similar to an idea proposed by Narayanan \etal~\cite{DBLP:conf/isqed/NarayananVPMV23}.
However, such structural properties always depend on the particular implementation, \ac{FPGA} architecture, surrounding logic, and synthesizer optimizations, hence calling for a more generic solution.

To address this issue, we developed a two-phased approach: 
We \circledtext[boxtype=O]{1}~identify candidates using architecture-dependent structural techniques and then \circledtext[boxtype=O]{2}~attempt to determine the functionality of each candidate using architecture-independent functional analysis.
We move the challenge of identifying operands and their bit order to the second stage, hence making it architecture agnostic.

\subsubsection*{\circledtext[boxtype=O]{1} Candidate Identification} 
Arithmetic operations are often implemented using carry gates for both Xilinx and Lattice \acp{FPGA}.
Because the characteristics of these carry gates can substantially differ between vendors, different approaches for structural candidate identification are required depending on the \ac{FPGA} architecture.
We \circledtext{1} start from specific structural patterns in the netlist, such as carry chains, and then \circledtext{2} generate the final candidates by additionally considering varying sets of preceding and succeeding combinational gates.
We do this because the neighboring combinational logic can differ vastly depending on the implemented arithmetic operation and the number of control inputs.

A salient advantage of this methodology is its generic nature, which relieves the user of the intricate task of manually crafting exact candidates.
So far, we implemented this structural approach for both Xilinx and Lattice iCE40 platforms.
To integrate other \ac{FPGA} architecture, the structural characteristics of that architecture must be investigated and techniques to compile structural candidates must be implemented.
This can be done in a matter of hours.

\subsubsection*{\circledtext[boxtype=O]{2} Functional Verification} 
Having identified structural candidates, we attempt to determine their implemented operations using functional methods to remain architecture-agnostic.  
We first determine bit order, operand memberships, and control signals for each structural candidate.
Afterwards, we check against a library of functional models of known arithmetic operations using an \ac{SMT} solver.

To this end, we first \circledtext{3} derive Boolean functions for all output nets from the sub-graph of each structural candidate.
Next, \circledtext{4} multiple different functional candidates are generated from the Boolean functions of each structural candidate to enable checking against different models from the library such as additions, subtractions, or comparisons.
Since the identification of operands, control inputs, and bit orders may be ambiguous, \circledtext{5} multiple variations of every functional candidate are created.
To reduce the number of variations, we leverage different functional characteristics depending on the operation to check for.
For example, for adder input bit orders, we consider how many output nets are influenced by each input signal.
The \acs{LSB} of an adder influences all of its outputs, while the \acs{MSB} usually influences only a single output.
Similarly, for an adder's output order, we consider how many inputs each output depends on.
Here, the \acs{LSB} depends on the least amount of inputs while the \acs{MSB} depends on the most.
Similar properties can be identified for other arithmetic operations.
For each functional candidate and its variations, \circledtext{6} we query an \ac{SMT} solver to identify the functionality of the structures given different assignments to their control values.

\subsubsection*{Evaluation}
We verified functionality under at least one control assignment for 94\% of the carry chains across our benchmarks, see \autoref{maggie23::tab::eval}.
Issues only arose when carry chains were strongly interwoven with surrounding control and data-path logic, which sometimes prevented the extraction of correct structural candidates.
Depending on the benchmark, we automatically verified that 1\% to 91\% (average: 32\%) of all combinational gates implement an arithmetic operation.
The variation in coverage is the result of different kinds of benchmark implementations.
For signal processing designs like the \texttt{canny\_edge\_detector} and the \texttt{fft64}, the data path mostly comprises of arithmetic operations.
\acs{CPU} implementations such as \texttt{ibex}, \texttt{icicle}, and \texttt{simple\_risc\_v} contain fewer carry chains and more control logic, which explains lower numbers in detected arithmetic operations and overall combinational coverage.

\subsection{Data Path}
\label{maggie23::subsec::data_path}
The idea behind the dataflow analysis methodology of \DANA~\cite{DBLP:journals/tches/AlbartusHTAP20} is, by nature, independent of the target gate type.
In our case study, we leveraged \DANA to identify registers and developed a separate but related technique to group \acs{MUX} gates into word-level structures.
Similar to \DANA, we used control signals as well as predecessor and successor gates to identify word-level \acsp{MUX}.
Accordingly, to generalize our data-path analysis, we extended \DANA to take on arbitrary, user-defined gate types, including \acsp{MUX}, instead of refining our initial approach.
In addition, we added the option to provide known word-level structures of any kind to \DANA to support word-level recovery of the target gate type.
For example, in our case study \DANA could only search for registers, and only previously identified registers could be provided as known groups.
Now, \DANA can also reconstruct word-level \acsp{MUX} while taking previously identified registers and arithmetic operations into account.

\subsubsection*{Evaluation}
To evaluate the success of register recovery, we followed the evaluation proposed for \DANA~\cite{DBLP:journals/tches/AlbartusHTAP20}.
We generated a ground truth by leveraging symbols left in the gate-level netlists of our benchmark by the synthesizer.
Then, we use \ac{NMI} and purity as metrics to compare our refined approach against the original \DANA implementation.
To this end, we fed known word-level structures such as \acp{BRAM}, \acp{DSP}, and arithmetic operations to \DANA as additional predefined knowledge to take into account during analysis.
This marginally improved \ac{NMI} and purity for most benchmarks, raising the \ac{NMI} by up to 0.14 and the purity by up to 0.30, see \autoref{maggie23::tab::eval}.

We also employed \DANA to recover word-level \acsp{MUX}.
However, by nature, no ground truth for \acsp{MUX} exists, as they are implicitly created from \ac{HDL} control flow structures by the synthesizer and then implemented in \acp{LUT} alongside other logic.
Therefore, it is, at best, ambiguous which \ac{HDL} constructs would be translated into a \ac{MUX}, and accurately measuring the quality of our results is impossible.
Still, we provide an intuition for their quality by investigating the sizes of recovered \acp{MUX} in \autoref{maggie23::app::mux_eval}.

\subsection{Bit Order}
\label{maggie23::subsec::bit_order}
Transferring our bit-order propagation algorithm from the case study to other \ac{FPGA} architectures was straightforward because the algorithm was already generic by nature.
It operates only on known pin groups of recovered structures and the netlist graph representation to determine successor and predecessor pin groups, which is applicable independent of the underlying gate types.
However, \acs{CPU} netlists tend to confuse our algorithm due to the interwoven data path of \acsp{ALU}.
To improve bit-order propagation results, we extended our approach for consensus-finding between conflicting bit orders propagated to the same pin group.
To this end, we now apply three consensus-finding mechanisms that try to recover a consecutive bit order.

Consider a module with a group of three input pins $[i_0, i_1, i_2]$.
After propagating pin indices from structures with known bit orders to those with unknown bit orders, each pin is annotated with multiple index candidates coming from different sources.
Here, $[0, 1, 2]$ is a list of indices gathered from the same source structure, denoting that $i_0$ should receive index $0$, $i_1$ index $1$, and $i_2$ index $2$.
We use $?$ to indicate missing index information for the respective pin and $X$ to show that the respective pin is no longer considered.

\subsubsection*{Shifted Consensus}
If conflicting pin orders are propagated to the pins of a pin group, but these orders are just shifted variants of each other, consensus is found and an order starting at index 0 is annotated.
For example, if orders $[1, 2, 3]$ and $[2, 3, 4]$ are propagated to a group of three pins, the pins will be assigned the order $[0, 1, 2]$.

\subsubsection*{Majority Consensus}
If the propagated pin orders of a pin group are in conflict, but a non-conflicting majority vote on each individual pin is feasible, the predominant bit order is annotated.
For example, if order $[0, 1, 2]$ is propagated to a group twice and order $[4, 3, 7]$ is propagated once, $[0, 1, 2]$ is annotated by majority decision.
Since index propagation is performed for each pin individually, there may not be a predominant index for every pin of a pin group.
If the majority vote fails for a single pin of a group, no index is annotated to any pin of that group.

\subsubsection*{Iterative Majority Consensus}
Sometimes, the same pin group index is propagated to multiple pins of the same group.
For example, the order $[0, ?, 2]$ may be propagated to a group twice and order $[?, 1, 1]$ once.
Here, the annotated indices $[?, 1, 1]$ will be ignored in the first iteration since they contain the same index for different pins of the same group, therefore conflicting with itself.
Instead, indices $0$ and $2$ will be annotated to the first and the last pin by majority vote.
In the second iteration, only the pins missing an index are considered.
The previously gathered groups of indices $[0, ?, 2]$ and $[?, 1, 1]$ are now transformed into $[X, ?, X]$ and $[X, 1, X]$, thereby resolving internal conflicts. 
Now $[X, ?, X]$ and $[X, 1, X]$ cause the second pin to be annotated with index $1$, finally resulting in $[0, 1, 2]$.

\subsubsection*{Evaluation}
To evaluate our bit-order propagation, we considered the bit orders given for \acsp{DSP} and \acsp{BRAM} and the ones recovered for operands of arithmetic operations to be correct and started propagation from there.
A reliable bit-order ground truth---given as labels left by the synthesizer---is only available for registers.
For evaluation, we created registers based on the ground truth used to evaluate \DANA.
Furthermore, we only consider registers containing more than three \acp{FF}, as smaller ones often belong to the control path.
In \autoref{maggie23::tab::eval}, we report the absolute number of pin groups we consider for bit-order propagation, the share of initially ordered pin groups, and the total share of ordered pin groups after propagation.
In addition, we report the proportion of ordered pin groups that are correct according to our ground truth.    
On average, we reconstructed 86\% of bit orders across our benchmarks with 97\% of them being correct, demonstrating the effectiveness of our method.
In line with expectations, we observed that the quality of bit-order propagation results improved with the number of structures featuring an inherently known bit order.

\subsection{Guided Symbolic Execution}
\label{maggie23::subsec::guided_symbolic}
Manually tracing signals across clock cycles in simulation to identify the sources of \acs{DSP} inputs in our case study turned out not to be scalable as it is tedious and error-prone, emphasizing the need for automation.
Hence, we implemented a \ac{SE} approach that is guided by concrete control values obtained from virtual probing or simulation to avoid state explosion.
This approach can be helpful beyond \ac{DSP} analysis to semi-automatically investigate a circuit's dynamic behavior.

Our approach allows symbolically evaluating any data signal and tracing it back to its origin to automatically generate equations describing that signal's behavior over time.
These equations depend only on previously defined endpoints such as global inputs, constants (\eg, from memory), or registers.
Thus, guided symbolic execution produces a system-level representation that abstracts away timing behavior and instead focuses on the sequence of computations. 
Finally, the human reverse engineer can manually simplify and interpret the extracted equations to assign high-level meaning.

Starting at the cycle of interest, our guided \ac{SE} approach first constructs the Boolean function describing the combinational sub-graph in front of a net.
Next, it replaces all previously identified control signals with concrete values from the simulation while keeping all other variables symbolic.
Each variable left in the Boolean function now corresponds to an output net of a sequential gate.
If the sequential gate is not one of our endpoints, it finds the last cycle in which the gate was updated and continues back-tracing from the data inputs of the gate at that cycle. 
This process is repeated until the resulting Boolean function only contains variables corresponding to the output nets of our endpoints.
To resolve potential recursive dependencies, our guided \ac{SE} approach introduces intermediate variables to break up these dependencies when they are detected.
To achieve this, we automatically identify sequential loops within the netlist and assign intermediate variables to the registers at the outputs of these loops.

\subsubsection*{Evaluation}
Applying guided \ac{SE} is highly specific to the target implementation and the goals of the reverse engineer.
Therefore, it is impossible to express its effectiveness in numbers.
Still, we applied guided \ac{SE} in a white-box case study and verified the results against a ground truth in \autoref{maggie23::subsec::white-box} to demonstrate correctness.
In a second evaluation step, we applied guided \ac{SE} on the Maggie netlist to highlight its efficiency and scalability.
To this end, we semi-automatically extracted a Python script describing the circuit's algorithmic behavior during execution after initially performing the same task manually in \autoref{maggie23::subsec::algorithmic_rec}.
Thereby, we reduced the required effort from multiple weeks to merely a matter of days.
    
\subsection{White-Box Case Study}
\label{maggie23::subsec::white-box}
To assess the applicability of our workflow, we conducted a white-box case study on a signal-processing design similar to Maggie.
This additional study is conducted because it enables a comparison between the recovered design and a known ground truth, which is impossible in the black box setting.
To this end, we synthesized an open-source Hilbert transformer design~\cite{hilbert_transformer_opencores} for a Xilinx 7-series \acs{FPGA} and then extracted an algorithmic abstraction from the resulting gate-level netlist using only our generalized techniques.

\paragraph{Word-Level Reconstruction}
The initial netlist comprised 425 combinational and 619 sequential gates.
After netlist pre-processing, we ended up with 1025 gates in total, 374 of them combinational and 651 sequential.
Next, we recovered word-level structures using our generalized algorithms.
The algorithm classified all 17 carry chains into 10 subtractions and 7 additions, identified 36 16-bit registers, and annotated 134 multi-bit pin groups to registers and arithmetic structures.
When comparing against the ground truth, \DANA achieved an \acs{NMI} of 1.00, and our bit-order propagation assigned a correct bitorder to all pin groups. 
In total, we automatically assigned over 95\% of all gates to word-level structures.

\paragraph{Algorithmic Recovery}
The Hilbert transformer comes with a testbench to validate its Verilog description before implementation.
The same testbench can also be used to simulate the gate-level netlist, thereby generating traces similar to those that we received from virtual probing in \autoref{maggie23::subsec::algorithmic_rec::virtual_probing}.
We successfully identified all relevant control signals in the netlist by searching for nets connecting to the control pins of registers, as this is a precondition to run guided \ac{SE} effectively.
Next, we fed concrete values generated through simulation with the provided testbench to our guided \ac{SE} algorithm for all identified control signals.
Finally, our guided \ac{SE} approach generated a word-level Boolean function for each global output that only depends on global inputs and the aforementioned intermediate variables introduced because of recursion.

\paragraph{High-Level Sensemaking}
We manually translated the recovered Boolean functions into Python code, which we again handed to a domain expert who was unaware of the nature of the analyzed design.
The expert identified the high-level functionality from the script and drew a block diagram that largely matched the one provided in the official documentation, see \autoref{maggie23::app::white-box} for a comparison.

Overall, our generalized techniques enabled us to recover a high-level functional description of the analyzed design in only a few days, a task that took months to complete for the iPhone netlist.
Still, significant manual effort is required despite this speedup.
Overall, this white-box case study allowed us to validate the results of our workflow against a ground truth, proving its effectiveness.
Furthermore, the verified success in this white-box case study supports the correctness of the analogous black-box case study, which cannot be directly verified in the same manner because the black-box setting precludes the existence of a known ground truth.

\begin{table*}[htb]
    \centering
    \caption{Results of our evaluation on the open-source benchmarks described in \autoref{maggie23::tab::benchmarks} and Maggie.}
    \label{maggie23::tab::eval}
    \resizebox{\textwidth}{!} {%
    \begin{tabular}{l l | R{0.4cm}R{0.4cm}R{0.4cm}R{0.4cm}R{0.4cm}R{0.4cm}R{0.4cm}R{0.6cm}R{0.7cm} | rrrr | rrrr }%
        \toprule%
        &\multicolumn{1}{c}{}& \mcrot{1}{l}{60}{\textbf{addition}} & \mcrot{1}{l}{60}{\textbf{subtraction}} & \mcrot{1}{l}{60}{\textbf{counter}} & \mcrot{1}{l}{60}{\textbf{negation}} & \mcrot{1}{l}{60}{\textbf{const. mul.}} & \mcrot{1}{l}{60}{\textbf{comparator}} & \mcrot{1}{l}{60}{\textbf{unknown}} & \mcrot{1}{l}{60}{\textbf{total}} & \mcrot{1}{l}{60}{\textbf{classified}} & \mcrot{1}{l}{60}{\textbf{NMI (org.)}} & \mcrot{1}{l}{60}{\textbf{NMI (our)}} & \mcrot{1}{l}{60}{\textbf{purity (org.)}} & \mcrot{1}{l}{60}{\textbf{purity (our)}} & \mcrot{1}{l}{60}{\textbf{no. of groups}} & \mcrot{1}{l}{60}{\textbf{initial ordered}} & \mcrot{1}{l}{60}{\textbf{final ordered}} & \mcrot{1}{l}{60}{\textbf{correct}} \\%
        \midrule%
        \textbf{Design}& \textbf{Vendor} & \multicolumn{9}{c|}{\textbf{Arithmetic}} & \multicolumn{4}{c|}{\textbf{\texttt{DANA}} (Registers)} & \multicolumn{4}{c}{\textbf{Bit Order}}\\
        \midrule
\midrule%
\multirow{2}{*}{\textbf{ibex}}&Lattice&5&0&4&0&0&0&0&9&0.05&0.98&0.98&0.97&0.96&168&0.27&0.40&1.00\\%
&Xilinx&1&0&2&0&0&0&3&6&0.01&0.98&0.98&0.96&0.97&148&0.16&0.89&1.00\\%
\midrule%
\multirow{2}{*}{\textbf{icicle}}&Lattice&2&0&7&0&0&1&0&10&0.12&0.86&0.92&0.89&0.87&112&0.31&0.62&0.94\\%
&Xilinx&1&1&4&0&0&0&1&7&0.05&0.91&0.94&0.88&0.95&346&0.79&0.87&1.00\\%
\midrule%
\multirow{2}{*}{\textbf{simple\_risc\_v}}&Lattice&3&1&4&0&0&6&0&14&0.13&0.99&0.98&0.98&0.96&139&0.33&0.95&1.00\\%
&Xilinx&3&1&4&0&0&2&0&10&0.07&0.98&0.98&0.99&0.99&117&0.29&0.91&1.00\\%
\midrule%
\multirow{2}{*}{\textbf{canny\_edge\_detector}}&Lattice&68&0&1&5&35&5&4&118&0.62&0.81&0.83&0.70&0.77&826&0.44&0.78&1.00\\%
&Xilinx&57&0&1&3&48&3&6&118&0.77&0.81&0.81&0.61&0.62&620&0.46&0.90&0.99\\%
\midrule%
\multirow{2}{*}{\textbf{fft64}}&Lattice&93&4&7&0&4&0&1&109&0.40&0.90&0.92&0.81&0.87&604&0.59&0.94&0.96\\%
&Xilinx&25&5&0&0&4&0&6&40&0.27&0.88&0.95&0.70&0.90&351&0.37&0.97&0.97\\%
\midrule%
\multirow{2}{*}{\textbf{sha256}}&Lattice&11&0&1&0&0&0&0&12&0.10&0.94&0.95&0.81&0.88&170&0.22&0.99&1.00\\%
&Xilinx&10&0&0&0&0&0&3&13&0.27&0.91&0.96&0.75&0.87&134&0.25&0.97&1.00\\%
\midrule
\midrule
\multirow{2}{*}{\textbf{hilbert\_transformer}}&Lattice&7&10&0&0&0&0&0&17&0.89&0.90&0.98&0.74&0.96&126&0.46&1.00&1.00\\%
&Xilinx&7&10&0&0&0&0&0&17&0.91&0.85&0.99&0.68&0.98&126&0.43&1.00&1.00\\%
\midrule%
\multirow{1}{*}{\textbf{maggie}}&Lattice&4&0&24&0&0&15&0&43&0.16&N/A&N/A&N/A&N/A&326&0.47&0.79&{N/A}\\%
        \bottomrule%
    \end{tabular}
}
\end{table*}
\section{Discussion}
\label{maggie23::sec::discussion}

\subsection{Revisiting Research Questions}

\label{maggie23::subsec::revisiting}
\subsubsection*{RQ1}
In our case study, we successfully recovered the algorithm controlling the Taptic Engine of the iPhone~7.
In response to \ref{maggie23::rq::1}, we now provide insights into the effort required to commit \ac{IP} theft.

While netlist recovery was solved following a known path with suitable open-source tooling, netlist analysis presented challenges that took months to complete.
In particular, the lack of hierarchy (\ref{maggie23::ch::1}), loss of data types (\ref{maggie23::ch::2}), and synthesizer optimizations (\ref{maggie23::ch::3}) initially hampered structural analysis, but could be dealt with once patterns were identified.
This was achieved through an interplay of manual and automated reverse engineering by first searching for repeating structures and then systematically automating the detection thereof.
While we automated substantial parts of the word-level reconstruction, manual inspection and correction were inevitably necessary to interpret results and address errors.

Classification of control logic (\ref{maggie23::ch::4}) had to be mostly done by hand as existing techniques provided limited insights, see \autoref{maggie23::subsec::word_level_rec}.
In the end, we bypassed in-depth analysis of \ac{FSM} state graphs and their interplay with the data path and other \acp{FSM} by relying on virtual probing.
Obtaining the \acs{IO} recordings required for this purpose presented engineering challenges on its own, see \autoref{maggie23::fig::pcb_probing}, but these are situational and depend on the \ac{PCB}.
The main challenge for recovering the implemented algorithm was understanding the dynamic circuit behavior (\ref{maggie23::ch::6}), which almost entirely depended on external data (\ref{maggie23::ch::5}).
To this end, simulation results were manually analyzed over many weeks.
Finally, understanding the implemented algorithm required domain knowledge beyond our reverse engineering expertise (\ref{maggie23::ch::7}) and called for a signal processing expert.

Throughout our case study, validating intermediate results within the black-box setting we were operating in presented another significant challenge. 
Subsequent stages always depended on previous findings, risking error propagation.
This concern was particularly pronounced during word-level reconstruction and algorithmic recovery, where our efforts often felt akin to groping in the dark.
Identifying and tracing errors was sometimes impossible.
While the \ac{SMT}-based verification of arithmetic operations offered an initial anchor point, this alone could not fully resolve the issue.
In this regard, insights derived from virtual probing played a crucial role in achieving end-to-end validation of our reverse engineering result.
Virtual probing was also instrumental in interpreting the data flowing through Maggie, enabling us to generate plots of processed data and extract equations representing the implemented algorithm.

An overshadowing issue of the entire case study was the lack of adequate, openly available tooling to automatically investigate the recovered netlist at scale.
Developing respective tooling, \eg, for netlist simulation, required immense effort and was the main reason for many of the months spent on conducting the case study.

\subsubsection*{RQ2}
In light of \ref{maggie23::rq::2}, we developed fully automated techniques that address many of the challenges we encountered in our case study.
Furthermore, we demonstrated their applicability across \ac{FPGA} architectures, particularly focusing on Lattice and Xilinx \acp{FPGA}.
We chose Xilinx as our second evaluation platform as they are the market leader among \ac{FPGA} vendors and their 7-series \acp{FPGA} exhibit higher complexity than the Lattice iCE40 device family.

Generalizing netlist pre-processing, data-path analysis, and bit-order reconstruction required only minor tweaks, as the techniques from our iPhone case study were already independent of the \acs{FPGA} architecture.
Here, we focused on addressing issues arising from different implementations ranging from \acsp{CPU} to \acs{DSP} designs.

While redesigning the functional arithmetic verification required significant effort and took months to complete, adding support for different architectures is now a matter of hours and only requires extending the structural candidate identification.
Furthermore, we showed that the labor-intensive dynamic extraction of an algorithmic description from Maggie can be sped up by our guided \ac{SE} approach.
This significantly reduced the time to get from word-level to an algorithmic description to at most a couple of days.

Still, we noticed that the quality of the results of our techniques varied depending on the \emph{kind} of implementation we analyzed.
\acs{DSP} designs similar to Maggie generally yield better results than \acsp{CPU}, presumably because the data and control paths in \acs{DSP} designs exhibit more structure.
As \ac{DSP} designs feature more arithmetic operations than, \eg, \acp{CPU}, we identify larger parts of their combinational data path.
\acsp{CPU} leverage varying bits of a register for a wide range of arithmetic operations and thus have a more interwoven data path, which sometimes required us to adapt our techniques.
In particular, introducing majority voting to the bit-order propagation improved results for \acs{CPU} implementations.

One challenge, however, remains unsolved: the identification and separation of control logic, see \autoref{maggie23::subsec::word_level_rec}.
We avoided this issue by using dynamic analysis, but it may not always be applicable, as dynamic analysis often requires observing an operational system.

By generalizing our techniques across different \ac{FPGA} architectures and implementations, we demonstrated that some of the netlist reverse engineering challenges we identified can be considered a one-time overhead.
The respective tooling only has to be developed once and can then be applied across a broad range of targets. 
Consequently, we contribute to significantly reducing the manual effort of \ac{FPGA} netlist reverse engineering by providing open-source implementations of the techniques presented in this section, allowing for a more realistic evaluation of countermeasures. 

\subsection{\acs{FPGA} Threats \& Defenses}
\label{maggie23::subsec::defenses}

\subsubsection*{Threats of \acs{FPGA} Reverse Engineering}
We showcased the feasibility of reverse engineering a real-world \ac{FPGA} design, even within the limitations of a resource-constrained academic setting.
Compared to regular \acp{IC}, the entry bar to reverse engineering is much lower.
Unlike \acp{FPGA}, for which reliable open-source bitstream documentation projects are readily available~\cite{xray_github,icestorm_github}, the extraction of a netlist from an \ac{IC} requires complex tooling, expensive equipment, and specialized expertise for preparation, imaging, and image analysis~\cite{DBLP:conf/aspdac/LippmannWUSEDGR19,DBLP:conf/ivsw/FyrbiakSKWERP17,DBLP:journals/jetc/QuadirCFASWCT16,DBLP:conf/dac/TorranceJ11,DBLP:conf/ches/TorranceJ09}.
Given these inherent challenges of \ac{IC} reverse engineering, we assume that \acp{IC} are more likely to be targeted by sophisticated attackers with nation-state-level resources.
In contrast, \acp{FPGA} are vulnerable to reverse engineering by less powerful adversaries.
On top of that, extracting an error-free netlist is much easier for \acp{FPGA} than it is for \acp{IC}, since for the latter, error-prone sample preparation and image analysis is needed while error-free bitstream documentation is more straight-forward to achieve.
A correctly recovered gate-level netlist enables techniques like virtual probing (and simulation in general) that are sensitive to errors in the netlist, thereby increasing the threat potential even further.

In addition to \ac{IP} theft, large-scale reverse engineering efforts can also provide an entry point for hardware Trojan insertion given the reconfigurability of \acp{FPGA}~\cite{DBLP:conf/woot/KatariaHPC19,DBLP:journals/jce/SwierczynskiFKM17,DBLP:journals/tcad/SwierczynskiFKP15}.
Consequently, a natural defense against netlist extraction would be to \textit{not} use \acp{FPGA} for sensitive \ac{IP} altogether.
We recognize that this is not a viable option in many cases, hence additional \ac{FPGA} countermeasures should be considered to defend against \ac{IP} theft and Trojan insertion.

\subsubsection*{Bitstream Protections}
A first line of defense is to impede the extraction of a gate-level netlist.
Proprietary bitstream formats may initially raise the bar for a reverse engineering attack, but given increasing automation, this may be overcome within a reasonable time.
This is inevitable since the user will always be able to translate \ac{HDL} designs into bitstreams using \ac{EDA} tools.
As offered by many \ac{FPGA} vendors, robust cryptographic schemes can ensure bitstream confidentiality, integrity, and authentication.
By encrypting and signing a bitstream, attacks such as \ac{IP} theft are impeded.
However, such bitstream protections are not always provided on low-cost \acp{FPGA} and legacy devices, which continue to be used for decades.
For example, the Lattice iCE40 \ac{FPGA} used in iPhone~7 does not feature any bitstream protections. 
Hence, Apple could not simply enable bitstream encryption to prevent reverse engineering attacks.

\subsubsection*{Netlist Protections}
In light of recent attacks on bitstream encryption schemes~\cite{DBLP:conf/uss/Ender0P20,DBLP:conf/ccs/MoradiBKP11,DBLP:conf/ctrsa/MoradiKP12,DBLP:conf/cosade/0001S16,DBLP:journals/trets/SwierczynskiMOP15,DBLP:conf/ccs/TajikLSB17}, further precautions must be taken to harden the netlist against static and dynamic analysis by deteriorating those netlist properties that facilitate automated reverse engineering.
Effective netlist obfuscation must aim at preventing the recovery of word-level structures as these structures are crucial for further analysis~\cite{DBLP:journals/corr/abs-2205-09892}.
Dynamic reverse engineering is key to achieve an algorithmic understanding of larger circuits.
Hence, to harden against such analysis techniques, non-simulatable primitives such as \acp{PUF}~\cite{DBLP:conf/iccad/WendtP14} or partial reconfiguration~\cite{DBLP:journals/tches/FyrbiakWDABTP18, DBLP:journals/tches/StolzASKNGFPGT21,DBLP:journals/tches/FyrbiakWDABTP18} should be included in the design and merged into the control and data paths. 

\subsection{Related Work}
\label{maggie23::subsec::related_work}
Existing reverse engineering techniques often only tackle isolated problems (such as bitstream reverse engineering or word-level reconstruction) and have been evaluated on (sometimes outdated) open-source benchmarks rather than black-box designs.
Hence, a comprehensive end-to-end real-world case study is still lacking in the literature.

\subsubsection*{FPGA Reverse Engineering}
When reverse engineering an \ac{FPGA}, the attacker commonly starts at the bitstream.
A comprehensive overview of the availability of bitstream encryption and known attacks thereon is given in \autoref{maggie23::app::bitstream_enc}.
The process of reverse engineering the proprietary bitstream format is well understood~\cite{DBLP:conf/aspdac/EnderSWWKP19,DBLP:conf/fpl/ZienerAT06,DBLP:conf/fpga/NoteR08,DBLP:conf/fpl/BenzSH12,DBLP:journals/mam/DingWZZ13,DBLP:conf/ets/ZhangTF23} and a variety of respective open-source tools exist~\cite{xray_github,DBLP:conf/fpl/KashaniEL22,debit,icestorm_github,DBLP:conf/date/PhamHK17}.
Ideally, such methods result in a database that allows the conversion of a bitstream into an error-free gate-level netlist describing the implementation on the analyzed \ac{FPGA}.
While we discuss existing netlist reverse engineering approaches below, none of the aforementioned works deals with the analysis of such an extracted gate-level netlist.
However, numerous publications have discussed how to meaningfully manipulate the bitstream without reverse engineering the netlist.
They often operate on the bitstream itself, \ie, without converting it to a netlist beforehand.
For example, Swierczynski \etal~\cite{DBLP:journals/tcad/SwierczynskiFKP15} demonstrate targeted bitstream manipulations to weaken cryptographic primitives and mount an attack on a real-world device~\cite{DBLP:journals/jce/SwierczynskiFKM17}.
Similarly, Moraitis \etal~\cite{DBLP:conf/date/MoraitisD20} attack an \ac{FPGA} implementation by replacing \ac{LUT} configuration strings.
In another case study, Kataria \etal~\cite{DBLP:conf/woot/KatariaHPC19} defeat Cisco router security measures by manipulating the bits corresponding to the \ac{FPGA} \ac{IO}.
Swierczynski \etal~\cite{DBLP:journals/tc/SwierczynskiBMP18} and Engels \etal~\cite{DBLP:conf/host/EngelsEP23} automate such manipulations and demonstrated that they (under certain conditions) can be mounted even on encrypted bitstreams.

\subsubsection*{Netlist Reverse Engineering}
For a comprehensive summary of prior work on netlist reverse engineering, see Azriel~\etal~\cite{DBLP:journals/jce/AzrielSAGMP21}.

One research strand focuses on automated state registers recognition and \ac{FSM} recovery~\cite{DBLP:conf/iscas/ShiTGR10,DBLP:conf/aspdac/MeadeZJ16,McElvain2001,DBLP:conf/iscas/MeadeJTZ16, DBLP:journals/jhss/MeadeSLDZJ18}.
\texttt{RELIC} by Meade~\etal~\cite{DBLP:conf/host/BrunnerBS19, DBLP:journals/jhss/MeadeSLDZJ18} rates the similarity of \ac{FF} fan-in trees to determine state and non-state \acp{FF}.
Chowdhury~\etal~\cite{DBLP:conf/iccad/ChowdhuryYN21} propose a \ac{GNN} base approach that can also separate control \acp{FF} from data.
To identify known sub-circuits, functional~\cite{DBLP:conf/host/LiWS12,  DBLP:conf/host/LiGSTTMSS13, DBLP:conf/date/SubramanyanTPRSM13, DBLP:journals/jhss/MeadeSLDZJ18} and graph-similarity based approaches~\cite{DBLP:journals/tc/FyrbiakWRBP20} have been proposed.
WordRev by Li~\etal~\cite{DBLP:conf/host/LiGSTTMSS13} comprises methods to recover word-level structures for functional matching.
Subramanyan~\etal~\cite{DBLP:conf/date/SubramanyanTPRSM13} builds on their work to identify components such as adders, multipliers, counters, and registers.
Albartus~\etal introduced \DANA~\cite{DBLP:journals/tches/AlbartusHTAP20} to identify high-level registers in a gate-level netlist.
Meade~\etal presented \texttt{REBUS} and \texttt{REWIND} to recover the datapath using similarity metrics comparable to \texttt{RELIC}~\cite{DBLP:journals/jhss/MeadeSLDZJ18}.
Alrahis~\etal~\cite{DBLP:journals/tcad/AlrahisSKPSMAS22} utilize \ac{GNN} to determine for each gate whether it belonged to a module that implements known functionality.
Narayanan \etal~\cite{DBLP:conf/isqed/NarayananVPMV23} also propose an identification of arithmetic operations by taking advantage of carry chains on \acp{FPGA}, as discussed in \autoref{maggie23::subsec::arithmetic}.
Some work has been done to partition a netlist based on the interconnectivity of a cluster of gates.
Werner~\etal~\cite{DBLP:conf/ivsw/WernerLBG18} evaluate the applicability of conventional graph clustering algorithms to circuits, while Hong~\etal~\cite{DBLP:journals/tai/HongLSG23} search for partitions by optimizing an n-cut with a \ac{GNN}.

\subsection{Limitations and Future Work}
\label{maggie23::subsec::limitations}
Our case study, while focused on a single \acs{DSP} design on a relatively small \ac{FPGA}, still provides valuable insights.
As we have shown in \autoref{maggie23::sec::generalization}, our case study findings inform generalized techniques that are applicable to implementations across different architectures, designs, and sizes.
Future research could build up on this work to explore more elaborate real-world designs on more complex \acp{FPGA} to further underline the threat potential of \ac{IP} theft for \acp{FPGA}.
Due to the very limited amount of resources available on Lattice iCE40 \acp{FPGA}, the benchmarks used for evaluation are quite small in terms of logic gates.
We are optimistic that our automated approaches scale well, even for larger netlists, while manual approaches naturally reach their limits rather quickly.
Our work builds the foundation for such case studies by providing the techniques to dissect such a netlist at scale.
Also, we have shown that there still is a need for better automated techniques for control logic extraction and analysis in gate-level netlists.

\section{Conclusion}
Our work highlights the challenges associated with \ac{IP} theft through \ac{FPGA} reverse engineering and discusses the issues arising in a real-world setting. 
To this end, our case study on a Lattice iCE40 \ac{FPGA} that is part of iPhone 7 underscores the vulnerability of \acp{FPGA} to \ac{IP} theft due to the low barrier for netlist extraction compared to regular \acp{IC}. 
Thereby, we reveal that while \ac{IP} theft from \acp{FPGA} requires substantial effort, specialized skills, and domain knowledge, most labor-intensive steps can be automated.
We contribute to the field by introducing generalized netlist reverse engineering techniques and open-source implementations, which reduce manual effort and facilitate future research in this domain. 
Our techniques prove effective across different \ac{FPGA} architectures and highlight the threat of \ac{IP} theft even by less resourceful adversaries.
This work emphasizes the importance of robust bitstream and netlist protections to safeguard valuable \ac{IP} in \acp{FPGA}. 
Finally, we call for more open and transparent research in \ac{FPGA} reverse engineering to better understand and mitigate associated threats.
\label{maggie23::sec::conclusion}

\begin{acks}
We thank Marc Fyrbiak and Max Hoffmann for the fruitful discussions.
Funded by the Deutsche Forschungsgemeinschaft (DFG, German Research Foundation) under Germany´s Excellence Strategy - EXC 2092 CASA - 390781972, through ERC grant 695022, NSF grants CNS-1563829 and CNS-1749845, and ISF project IZ25-5793-2019-43 (\textit{CHIoSec}).
\end{acks}

\bibliographystyle{ACM-Reference-Format}
\bibliography{bibliography}

\appendix

\section{Simulation in \HAL}
\label{maggie23::app::hal_simulation}
\begin{figure}[!htb]
  \centering
  \includegraphics[width=1\linewidth]{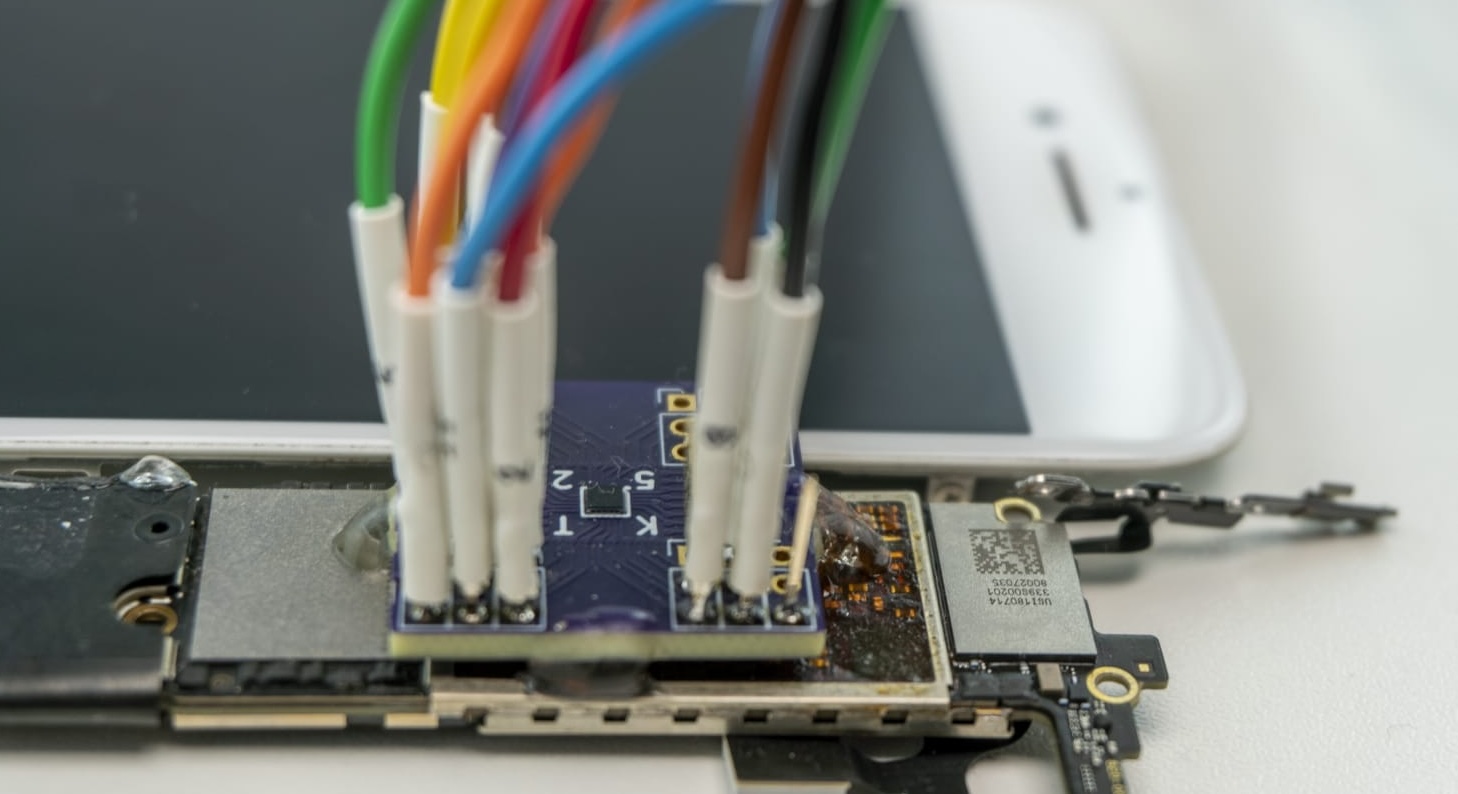}
  \caption{Side view of our custom breakout board sitting between the iPhone \ac{PCB} and Maggie.}
  \label{maggie23::fig::probing_setup_side}
\end{figure}

\begin{figure}[!htb]
  \centering
  \includegraphics[width=1\linewidth]{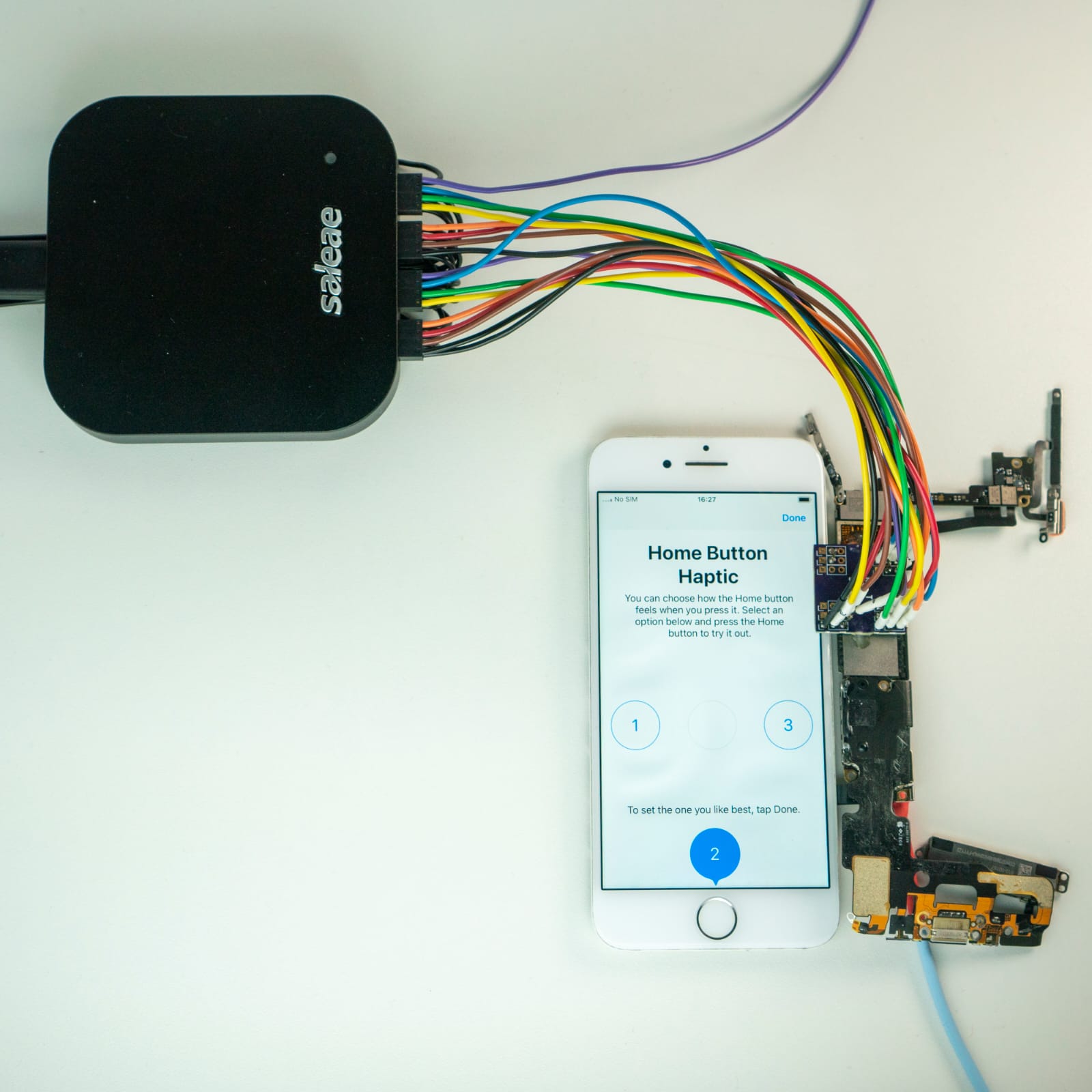}
  \caption{Top view of a Saleae logic analyzer being connected to the iPhone via our custom breakout board.}
  \label{maggie23::fig::probing_setup_top}
\end{figure}

\begin{figure*}[!hb]
  \centering
  \includegraphics[width=1\linewidth]{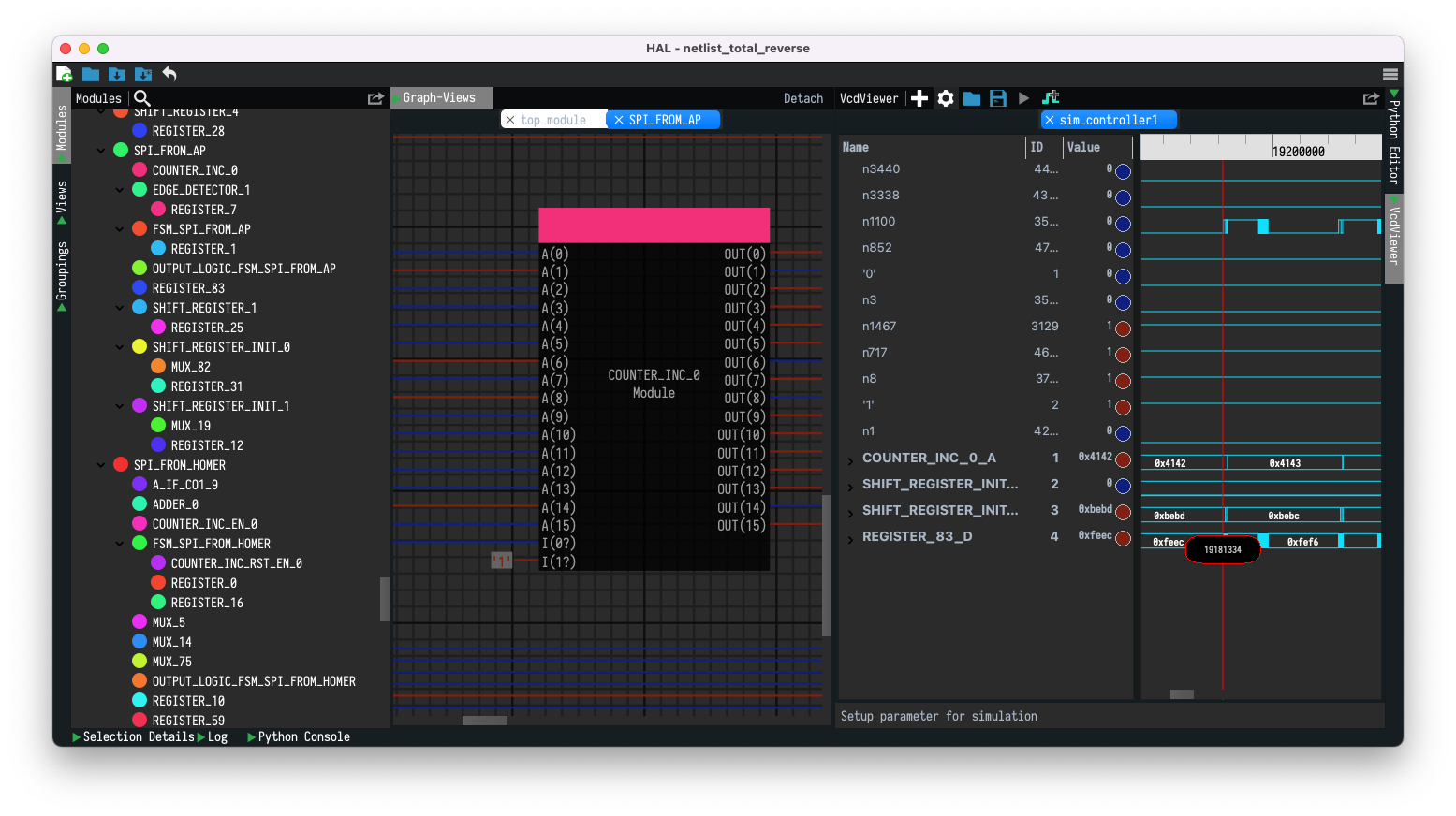}
  \caption{Screenshot of \HAL showing the module tree on the left, the graph view in the middle, and the simulator on the right. 
  The nets are colored depending on their current value, blue represents a \texttt{0} and red a \texttt{1}.}
  \label{maggie23::fig::hal_virtual_probing_screenshot}
\end{figure*}

For netlist simulation in \HAL, \eg, in the context of virtual probing, we extended the framework with an interface for commercial-grade simulation tools such as Verilator~\cite{verilator_github}.
Furthermore, we added a waveform viewer to the \HAL \acs{GUI} that can be controlled via a Python \acs{API}, see \autoref{maggie23::fig::hal_virtual_probing_screenshot}.
The dynamic state of the netlist can also be represented by coloring the nets in the graph view according to their current signal value, which adapts whenever the cursor in the waveform viewer is moved.
See \autoref{maggie23::fig::probing_setup_side} and \autoref{maggie23::fig::probing_setup_top} for pictures of our actual virtual probing setup as initially described in \autoref{maggie23::fig::pcb_probing}.

\section{Details on the Maggie DSP}
\label{maggie23::app::dsp_details}

In the following, we briefly comment on three key building blocks of the \ac{DSP}~implementation, as shown in~\autoref{maggie23::fig::maggie_dsp_overview}: 
$(i)$~an input processing, 
$(ii)$~a state observer to track the actuator dynamics, and 
$(iii)$~a closed-loop controller making the actuator follow a desired movement.

\subsubsection*{Input Processing}
Maggie's data path takes two input signals supplied via \acs{I2S} and \acs{SPI} from the motor driver and Homer, respectively. 
These two signals are combined (scaled and subtracted) and then passed to a look-up table-based function where linear interpolation is used to enhance resolution.
We believe that this step translates sensor measurements to the current \ac{LRA} position, relying on calibration.

\subsubsection*{State Observer}
The input processing result and the closed-loop controller output are passed to a processing stage that implements a state observer, estimating the \ac{LRA}'s position and velocity. In particular, Maggie appears to implement a \ac{PI}~observer~\cite{beale1989_piobserver}.

\subsubsection*{Closed-Loop Controller}
The closed-loop controller takes a time series of desired actuator positions as its reference signal. This signal determines the haptic event to be generated and, therefore, is reconfigurable. 
For instance, it is updated with the haptic feedback setting visible in \autoref{maggie23::fig::probing_setup_top}. 
Using the reference signal and the state observer outputs, the closed-loop controller accumulates the position and the velocity error (the desired velocity is obtained from numerical differentiation of the reference signal) and subtracts the integrated position estimation error.
The particular combination of error components suggests that the control loop is a \ac{PID}-controller~\cite{ellis2002observers, ellis2012control,  wu2017_servoperformance}. 
Before being passed to the motor driver via~\acs{I2S}, a low-pass filter is applied to the output of the closed-loop controller. 
We also obtained specifics of the implementation such as the concrete \ac{PID} values, \ac{IIR} filter coefficients, and initialization values. However, to protect Apple's \ac{IP}, we do not disclose such details.
\section{Benchmarks}
\label{maggie23::app::benchmarks}
\autoref{maggie23::tab::benchmarks} depicts the resource utilization of our six benchmark designs.
Differences in resource utilization between the Xilinx and Lattice benchmarks, \eg, for \texttt{fft64}, can be the result of the Xilinx synthesizer using \acsp{DSP} where the Lattice synthesizer resorts to carry chains.
Furthermore, Xilinx \acp{FPGA} leverage 6-input \acp{LUT} while Lattice \acp{FPGA} use 4-input ones.

\begin{table}[htb]
    \centering
    \caption{Open-source benchmark resource utilization in the number of gates.}
    \label{maggie23::tab::benchmarks}
    \resizebox{.99\linewidth}{!}{
    \begin{tabular}{l l | rrrrrr | r }%
        \toprule%
        \textbf{Design}&  \textbf{Vendor} & \textbf{LUT} & \textbf{Carry} & \textbf{FF} & \textbf{BRAM} & \textbf{DSP} & \textbf{Other} & \textbf{Total}\\
        \midrule
        \midrule%
        \multirow{2}{*}{\textbf{ibex~\cite{ibex_github}}}
        &Lattice&6379&295&1963&16&0&78&8731\\%
        &Xilinx&3503&65&1937&0&1&727&6233\\%
        \midrule%
        \multirow{2}{*}{\textbf{icicle~\cite{icicle_github}}}
        &Lattice&2705&262&983&20&0&59&4029\\%
        &Xilinx&1525&65&1009&270&0&78&2947\\%
        \midrule%
        \multirow{2}{*}{\textbf{simple\_risc\_v~\cite{simple_risc_v_github}}}
        &Lattice&2876&274&1235&6&0&31&4422\\%
        &Xilinx&1617&76&1003&6&0&37&2739\\%
        \midrule%
        \multirow{2}{*}{\textbf{canny\_edge\_detector~\cite{canny_edge_detector_opencores}}}
        &Lattice&3298&2170&4484&32&0&68&10052\\%
        &Xilinx&2538&636&3713&24&0&72&6983\\%
        \midrule%
        \multirow{2}{*}{\textbf{fft64~\cite{fft64_github}}}
        &Lattice&5362&2008&2556&14&0&124&10064\\%
        &Xilinx&1722&212&1988&3&4&160&4089\\%
        \midrule%
        \multirow{2}{*}{\textbf{sha256~\cite{sha256_secworks}}}
        &Lattice&3935&346&2378&0&0&90&6749\\%
        &Xilinx&2196&104&1830&0&0&269&4399\\%
        \midrule
        \midrule
        \multirow{2}{*}{\textbf{hilbert\_transformer~\cite{hilbert_transformer_opencores}}}
        &Lattice&445&225&569&2&0&78&1349\\%
        &Xilinx&304&68&587&0&0&85&1044\\%
        \midrule%
        \textbf{maggie}&Lattice&3241&422&1342&18&4&19&5046\\%
        \bottomrule%
    \end{tabular}
    }
\end{table}
\section{MUX Evaluation}
\label{maggie23::app::mux_eval}
\begin{figure*}
    \centering
    \begin{subfigure}[b]{0.49\textwidth}
        \centering
        \includegraphics[width=\textwidth]{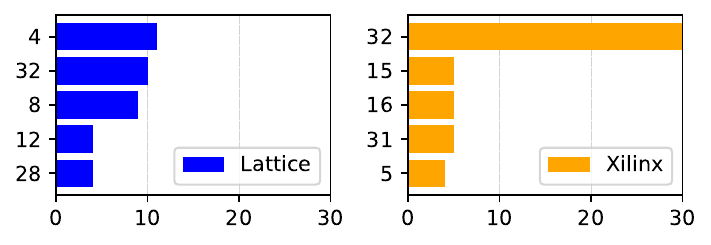}
        \caption{\texttt{ibex} (Expected: 32)}
        \label{maggie23::fig::muxer::ibex}
    \end{subfigure}
    \hfill     
    \begin{subfigure}[b]{0.49\textwidth}
        \centering
        \includegraphics[width=\textwidth]{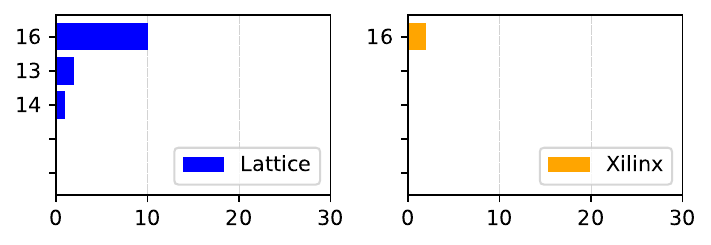}
        \caption{\texttt{canny\_edge} (Expected: 8,16,32)}
        \label{maggie23::fig::muxer::canny_edge_detector}
    \end{subfigure}
    \hfill
    \begin{subfigure}[b]{0.49\textwidth}
        \centering
        \includegraphics[width=\textwidth]{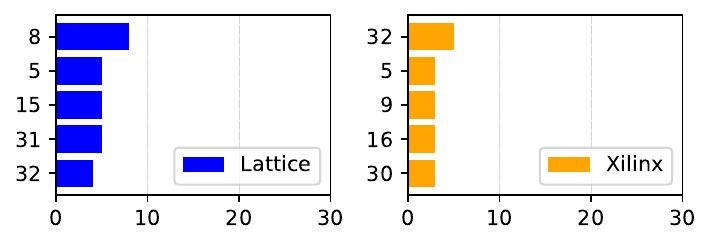}
        \caption{\texttt{icicle} (Expected: 32)}
        \label{maggie23::fig::muxer::icicle}
    \end{subfigure}
    \hfill
    \begin{subfigure}[b]{0.49\textwidth}
        \centering
        \includegraphics[width=\textwidth]{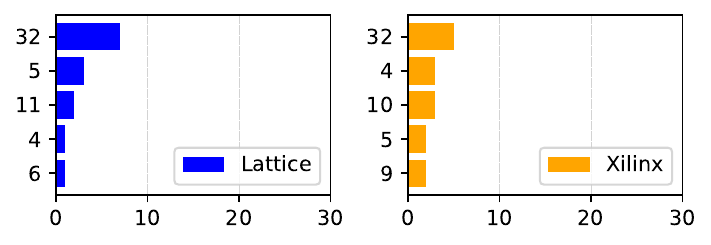}
        \caption{\texttt{simple\_risc\_v} (Expected: 32)}
        \label{maggie23::fig::muxer::simple_risc_v}
    \end{subfigure}
    \hfill
    \begin{subfigure}[b]{0.49\textwidth}
        \centering
        \includegraphics[width=\textwidth]{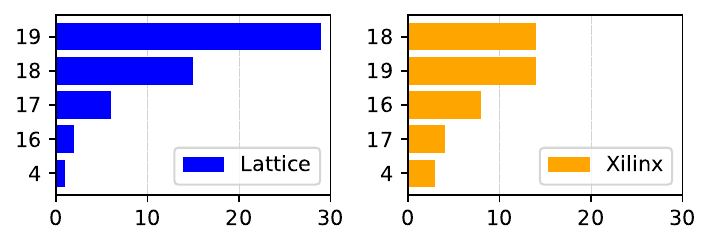}
        \caption{\texttt{fft64} (Expected: 19,18,17,16)}
        \label{maggie23::fig::muxer::fft64}
    \end{subfigure}
    \hfill
    \begin{subfigure}[b]{0.49\textwidth}
        \centering
        \includegraphics[width=\textwidth]{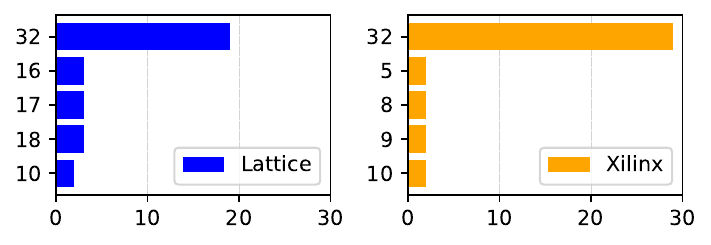}
        \caption{\texttt{sha256} (Expected: 256,32)}
        \label{maggie23::fig::muxer::sha256}
    \end{subfigure}
    \begin{subfigure}[b]{0.49\textwidth}
        \centering
        \includegraphics[width=0.53\textwidth]{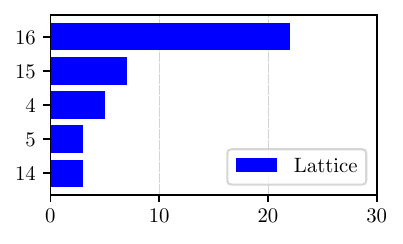}
        \caption{Maggie (Expected: 16)}
        \label{maggie23::fig::muxer::maggie}
    \end{subfigure}
    \caption{Evaluation of our word-level \acs{MUX} detection using \DANA.}
    \label{maggie23::fig::muxer}
\end{figure*}

\acsp{MUX} of sizes that divide the bit-size of the datapath or are multiples of a byte are likely to be correct. 
To this end, we plotted the distributions of the top 5 \acs{MUX} sizes for all benchmarks in \autoref{maggie23::fig::muxer}.
The figure illustrates the five most frequently recovered \acs{MUX} sizes for each benchmark, once synthesized for a Xilinx 7-series \acs{FPGA} and once for s Lattice iCE40 \acs{FPGA}.
In this analysis, we exclusively consider \acsp{MUX} comprising at least four gates. 
For the \texttt{hilbert\_transformer}, no word-level \acp{MUX} were identified, hence we omitted it in \autoref{maggie23::fig::muxer}.
We also provide our results for Maggie, which are only available for Lattice.

\section{White-Box Case Study Comparison}
\label{maggie23::app::white-box}

\begin{figure*}[!hb]
  \centering
    \centering
    \begin{subfigure}[b]{1\textwidth}
        \centering
        \includegraphics[width=0.8\linewidth]{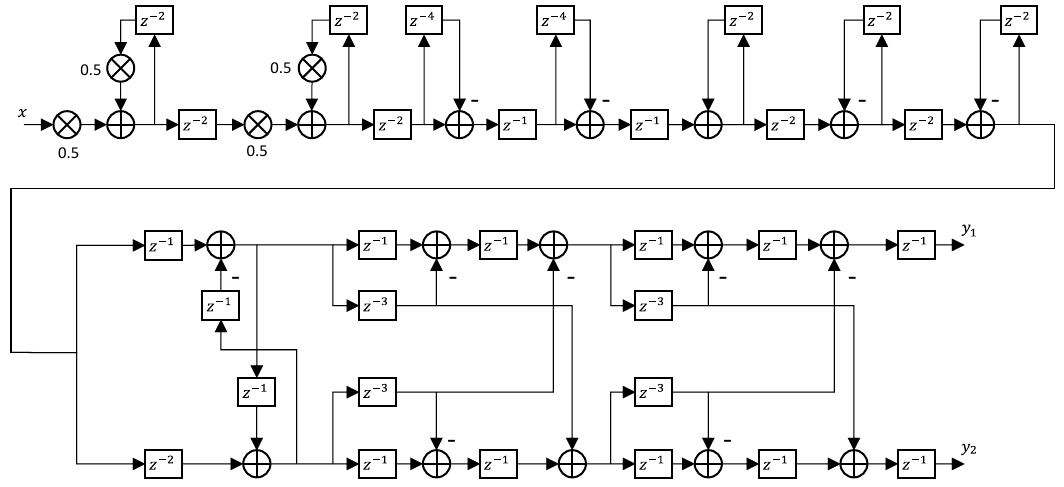}
        \caption{Block diagram based on recovered Boolean functions.}
        \label{maggie23::fig::hilbert_extract}
    \end{subfigure}
    \vfill
    \begin{subfigure}[b]{1\textwidth}
        \centering
        \includegraphics[width=0.85\linewidth]{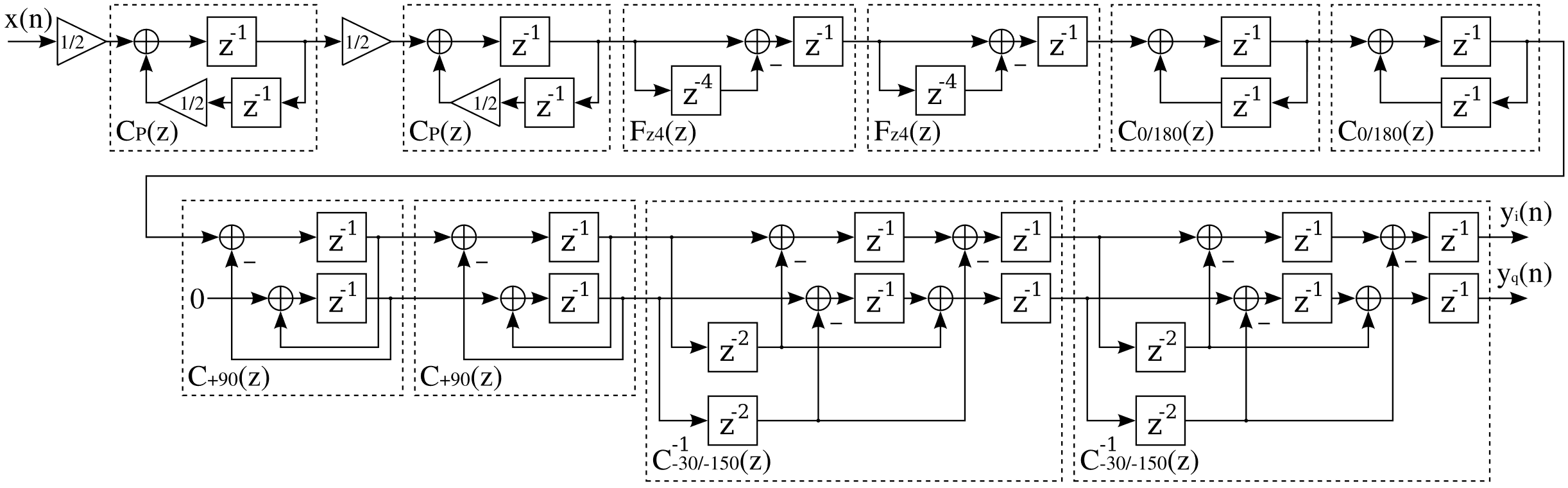}
        \caption{Block diagram from official documentation~\cite{hilbert_transformer_opencores, hilbert_transformer_documentation}.}
        \label{maggie23::fig::hilbert_official}
    \end{subfigure}
  
  \caption{Comparison of Hilbert transformer block diagrams in the white-box case study.}
  \label{maggie23::fig::hilbert_block_diags}
\end{figure*}

In our white-box case study, we used an open-source Hilbert transformer design~\cite{hilbert_transformer_opencores} as the ground truth to validate our generalized techniques. 
A signal processing expert unaware of the nature of the analyzed design inspected the Python code representing the recovered Boolean functions to draw a block diagram.
From their analysis, they correctly identified the Hilbert transformer. 
\autoref{maggie23::fig::hilbert_block_diags} shows the block diagrams representing the implemented signal processing chain. 
The diagram recovered from the gate-level netlist is shown in \autoref{maggie23::fig::hilbert_extract} and the one taken from the implementation's documentation~\cite{hilbert_transformer_documentation} in \autoref{maggie23::fig::hilbert_official}. 
While the recovered block diagram largely resembles the one from the documentation, partially matching exactly, there are also notable differences. 
Some are due to visualization choices, \eg, in favor of symmetry in the lower signal path, and others are due to implementation details affecting intermediate signals.
For example, delay elements (represented by negative powers of~$z$) can easily be moved around by factoring them into or out of parts of the processing. Similarly, delay elements that simply shift the entire output signal may be added or removed without altering the time-invariant processing behavior. 
Furthermore, the trivial addition with $0$ denoted in the documentation (first dashed box labeled $C_{+90}(z)$ in~\autoref{maggie23::fig::hilbert_official}) is not found in the netlist, likely due to synthesizer optimizations.
Please note that, despite the deviation of the visual representations and minor trivial differences, both block diagrams still describe the same overall processing. 

\section{Bitstream Encryption in the Wild}
\label{maggie23::app::bitstream_enc}
Here we provide an overview of the availability of bitstream encryption on many modern \acp{FPGA} from the top four \ac{FPGA} vendors, namely AMD Xilinx, Intel Altera, Lattice Semiconductor, and Microchip.
This list has been compiled from vendor documentation and to the best of our knowledge, but such resources may be incomplete or lack sufficient detail. Hence, we cannot guarantee correctness or completeness.
Generally, older and cost-optimized \acp{FPGA} can be seen to lack bitstream encryption, while modern high-end \acp{FPGA} usually offer respective protections.
While this list gives an intuition of how often vendors offer such features, we cannot give any numbers for how often these protections are actually enabled in the field.
Historically, Xilinx \acp{FPGA} have received more attention from the academic community due to their market-leader position and availability to researchers, resulting in more successful attacks on their encryption schemes compared to other vendors.
Hence, the lack of attacks on other vendors does not mean that their \acp{FPGA} are more secure when facing real-world attacks.
All listed attacks strictly require access to the \ac{FPGA} hardware.

\begin{table}[htb]
    \caption{Overview of the availability of bitstream encryption across different vendors and \ac{FPGA} families. We only state whether bitstream encryption is generally available for all~(\cmark), some~(\dmark), or none~(\xmark) of the \acp{FPGA} of a family, and whether attacks on the bitstream encryption have been demonstrated in academic literature. However, we do not evaluate the cryptographic strength of the encryption schemes.}
    \label{maggie23::tab::benchmarks}
    \resizebox{\linewidth}{!}{%
    \begin{tabular}{l | l c c | l c c}%
        \toprule%
        & \textbf{Family} & \textbf{Enc?} & \textbf{Attacks} & \textbf{Family} & \textbf{Enc?} & \textbf{Attacks} \\ 
        \midrule
        \midrule%
        \multirow{6}{*}{\rotatebox{90}{\textbf{AMD Xilinx}}} & \textbf{Virtex II} & \dmark & \cite{DBLP:conf/ccs/MoradiBKP11} & \textbf{Spartan 6} & \dmark & \cite{DBLP:journals/iacr/MoradiKP11,DBLP:conf/cosade/0001S16} \\
                                 & \textbf{Virtex 4} & \cmark & \cite{DBLP:journals/iacr/MoradiKP11,DBLP:conf/ctrsa/MoradiKP12,DBLP:conf/cosade/0001S16} & \textbf{Spartan 7} & \dmark & \cite{DBLP:conf/uss/Ender0P20,DBLP:conf/cosade/0001S16} \\
                                 & \textbf{Virtex 5} & \cmark & \cite{DBLP:journals/iacr/MoradiKP11,DBLP:conf/ctrsa/MoradiKP12,DBLP:conf/cosade/0001S16} & \textbf{Artix 7} & \cmark & \cite{DBLP:conf/cosade/0001S16,DBLP:conf/uss/Ender0P20,DBLP:conf/cosade/0001S16} \\
                                 & \textbf{Virtex 6} & \cmark & \cite{DBLP:conf/uss/Ender0P20,DBLP:conf/cosade/0001S16} & \textbf{Kintex 7} & \cmark & \cite{DBLP:conf/cosade/0001S16,DBLP:conf/uss/Ender0P20,DBLP:conf/ccs/TajikLSB17} \\
                                 & \textbf{Virtex 7} & \cmark & \cite{DBLP:conf/uss/Ender0P20,DBLP:conf/cosade/0001S16} & \textbf{Ultrascale} (all) & \cmark & \cite{DBLP:conf/fccm/EnderLMP22,DBLP:journals/tches/LohrkeTKBS18} \\
                                 & \textbf{Spartan 3} & \xmark & - & \textbf{Ultrascale+} (all) & \cmark & \cite{DBLP:conf/fccm/EnderLMP22} \\
        \midrule
        \multirow{6}{*}{\rotatebox{90}{\textbf{Intel Altera}}} & \textbf{Agilex} (all) & \cmark & - & \textbf{Stratix IV} & \cmark & - \\
                                 & \textbf{Arria} (all) & \cmark & - & \textbf{Stratix V} & \cmark & - \\
                                 & \textbf{Max V} & \xmark & - & \textbf{Stratix 10} & \cmark & - \\
                                 & \textbf{Max 10} & \cmark & - & \textbf{Cyclone IV} & \xmark & - \\
                                 & \textbf{Stratix II} & \cmark & \cite{DBLP:conf/fpga/MoradiOPS13,DBLP:journals/trets/SwierczynskiMOP15} & \textbf{Cyclone V} & \cmark & - \\
                                 & \textbf{Stratix III} & \cmark & \cite{DBLP:journals/trets/SwierczynskiMOP15} & \textbf{Cyclone 10} & \cmark & - \\
        \midrule
        \multirow{7}{*}{\rotatebox{90}{\textbf{Lattice}}} & \textbf{iCE40} & \xmark & - & \textbf{CertusPro NX} & \cmark & - \\
                                 & \textbf{ECP2} & \dmark & - & \textbf{Avant} & \dmark & - \\
                                 & \textbf{ECP3} & \cmark & - & \textbf{Mach NX} & \cmark & - \\
                                 & \textbf{ECP5} & \cmark & - & \textbf{MachXO2} & \xmark & - \\
                                 & \textbf{CrossLink} & \xmark & - & \textbf{MachXO3} & \xmark & - \\
                                 & \textbf{CrossLink NX} & \cmark & - & \textbf{MachXO5} & \cmark & - \\
                                 & \textbf{Certus NX} & \cmark & - & \textbf{XP2} & \cmark & - \\
        \midrule
        \multirow{4}{*}{\rotatebox{90}{\textbf{Microchip}}} & \textbf{ProASIC} & \xmark & - & \textbf{IGLOO 2} & \cmark & - \\
                                 & \textbf{ProASIC 3} & \cmark & \cite{DBLP:conf/ches/SkorobogatovW12} & \textbf{Fusion} & \cmark & - \\
                                 & \textbf{PolarFire} & \cmark & - & \textbf{SmartFusion} & \cmark & - \\
                                 & \textbf{IGLOO} & \cmark & - & \textbf{SmartFusion 2} & \cmark & - \\
        \bottomrule
    \end{tabular}
    }
\end{table}

\end{document}